\documentclass[aps,prb,amsmath,amssymb,showpacs,twocolumn,superscriptaddress]{revtex4-2}
\usepackage{amsmath}
\usepackage{amssymb}
\usepackage{amsthm}
\usepackage{setspace}
\usepackage{graphicx}
\usepackage{braket}
\usepackage{mathrsfs}
\usepackage[none]{hyphenat}
\usepackage{float}
\usepackage[colorlinks = true,linkcolor = blue,urlcolor  = blue,citecolor = blue,anchorcolor = blue]{hyperref}
\usepackage[utf8]{inputenc}
\usepackage[english]{babel}
\usepackage{bm}
\usepackage{csquotes}
\usepackage{color}
\usepackage[normalem]{ulem} 

\begin{document}
\title{\textcolor{black}{Semiclassical magnetotransport in Weyl semimetals with skyrmion-induced emergent magnetic fields}}

\author{Azaz Ahmad}
\affiliation{Department of Physics, Indian Institute of Technology, Kanpur-208016, India.}
\affiliation{School of Physical Sciences, Indian Institute of Technology Mandi, Mandi 175005, India.}
\affiliation{Department of Applied Physics, Tokyo University of Science, Katsushika, Tokyo 125-8585, Japan.}
\author{Takami Tohyama}
\affiliation{Department of Applied Physics, Tokyo University of Science, Katsushika, Tokyo 125-8585, Japan.}

\date{\today}

\begin{abstract}
We investigate magnetotransport in a time-reversal symmetry–broken,
untilted Weyl semimetal in the simultaneous presence of momentum-space
Berry curvature and \textcolor{black}{skyrmion-induced emergent magnetic fields $\mathbf{B}_{\mathrm{emer}}$}. Using a semiclassical
Boltzmann approach incorporating Berry-curvature corrections and
intervalley scattering, we analyze the longitudinal
magnetoconductivity and planar Hall conductivity in this mixed-topology
regime. In the absence of $\mathbf{B}_{\mathrm{emer}}$, increasing intervalley
scattering drives a strong sign reversal of the longitudinal
magnetoconductivity. A finite $\mathbf{B}_{\mathrm{emer}}$ introduces an additional
shift of the parabolic magnetic-field dependence, leading to a weak
sign-reversal regime without altering the curvature. The coexistence of
these effects naturally produces a strong-and-weak sign-reversal regime,
demonstrating that intervalley scattering and real-space topology control
distinct geometric features of the response. The emergent field further
induces asymmetry in the angular dependence of both longitudinal and
planar Hall conductivities. We show that a finite planar Hall response
can arise solely from $\mathbf{B}_{\mathrm{emer}}$ when its direction is varied,
providing a transport signature of real-space topology. Our results
establish that the skyrmion-induced emergent field acts as an independent
topological tuning parameter, revealing measurable consequences of the
interplay between real- and momentum-space Berry curvature in Weyl
systems.
\end{abstract}

\maketitle
\section{Introduction}
\label{sec:introduction}
Topological semimetals realized in kagome-lattice materials provide a natural platform in which distinct forms of topology can coexist and mutually influence electronic transport~\cite{le2026ultra,dominici2023coupled,he2024enhanced,raju2024anisotropic}. In Weyl semimetals (WSMs), momentum $\mathbf{k}$-space topology arises from isolated band-touching points (Weyl nodes) that act as monopoles of Berry curvature $\boldsymbol{\Omega}^{\chi}_{\mathbf{k}}$ with chirality
$\chi=\pm 1$~\cite{wan2011topological,armitage2018weyl}. The associated Berry curvature profoundly modifies semiclassical transport, generating anomalous velocity contributions and phase-space corrections that underlie phenomena such as the anomalous Hall effect and the chiral anomaly
\cite{son2012berry,burkov2014anomalous,bevan1997momentum,volovik1999induced,volovik2003universe,novoselov2006unconventional}. These effects are now well
established in Weyl systems~\cite{berry1984quantal,
sundaram1999wave,xiao2010berry,nagaosa2013topological,verma2022unified,goswami2013axionic}.

Kagome-lattice materials, however, offer an additional ingredient:
nontrivial real-space spin textures. Owing to geometrical frustration,
strong spin–orbit coupling, and magnetic ordering tendencies, kagome systems
can host complex magnetic configurations including noncoplanar spin states
and skyrmion textures~\cite{hirschberger2019skyrmion,hirschberger2024lattice,du2020room}.
When the electronic spins adiabatically follow such a
smoothly varying magnetization texture $\mathbf{n}(\mathbf{r})$, they acquire
a real-space Berry phase that manifests as an emergent magnetic field $\mathbf{B}_{\mathrm{emer}} = \frac{\hbar}{2e}\mathbf{n}\cdot(\partial_x \mathbf{n}\times\partial_y \mathbf{n}) ~\hat{e}$, where $~\hat{e}$ is the unit vector pointing in the direction of the field. ${\mathbf{B}_\mathrm{emer}}$ acts on the charge carriers analogously to an external magnetic field $\mathbf{B}${~\cite{nagaosa2013topological}}. This mechanism is responsible for the topological Hall effect in chiral magnets and has been widely discussed in systems with noncoplanar spin order~\cite{
nagaosa2013topological,vir2019anisotropic,addison2025anomalous,gobel2025topological}. \\

In a kagome WSM hosting both Weyl nodes and a nontrivial magnetic
texture, electrons therefore experience two distinct topological fields:
the Berry curvature in momentum space and the emergent magnetic field in
real space~\cite{le2026ultra,dominici2023coupled,he2024enhanced,raju2024anisotropic}. Within the semiclassical Boltzmann framework, the equations of
motion depend only on the total magnetic field
$\mathbf{B}_{\mathrm{tot}}=\mathbf{B}+\mathbf{B}_{\mathrm{emer}}$. As a result, the modified
phase-space factor that controls the dynamics of the electrons becomes $\mathcal{D}^{\chi} = \left(1 + \frac{e}{\hbar}
\mathbf{B}_{\mathrm{tot}}\cdot
\boldsymbol{\Omega}^{\chi}_{\mathbf{k}}\right)^{-1}$\cite{son2012berry,ahmad2021longitudinal}. This expression explicitly contains the cross term
$\mathbf{B}_{\mathrm{emer}}\cdot
\boldsymbol{\Omega}^{\chi}_{\mathbf{k}}$, which encodes the
interplay between real-space and momentum-space topology.
Importantly, the intrinsic Berry curvature of the Weyl nodes remains
unchanged; rather, it is the emergent field that modifies the effective
magnetic environment in which the Weyl quasiparticles propagate. Consequently, magnetotransport in such systems carries signatures of a mixed real–momentum topological response. In particular, longitudinal magnetoconductivity (LMC) and planar Hall conductance (PHC) acquire corrections controlled by $\mathbf{B}_\mathrm{emer}$, allowing magnetic texture effects to directly influence anomaly-driven transport. Kagome Weyl materials therefore provide an experimentally relevant setting in which this real–momentum topological mixing can be probed through standard magnetotransport measurements. A central conceptual advance of this work is the identification of a
\emph{mixed topological magnetotransport mechanism} arising from the interplay between real-space and momentum-space topology. While previous studies have focused on strain-induced axial fields
that are introduced as chirality-dependent gauge fields by construction ~\cite{ahmad2023longitudinal,goswami2013axionic,volovik1999induced,liu2013chiral,grushin2012consequences,zyuzin2012topological},
here the chirality dependence of the transport response emerges
\emph{dynamically} from the coupling between the real-space emergent
field and the intrinsic Berry curvature of Weyl fermions.\\

\textcolor{black}{Although the emergent magnetic field $\mathbf{B}_{\rm emer}$ enters the semiclassical equations of motion in a form analogous to an external magnetic field, its physical origin is fundamentally different. In particular, $\mathbf{B}_{\rm emer}$ is generated by the scalar spin chirality associated with the underlying real-space magnetic texture and is therefore directly tied to the skyrmion density and chirality. Consequently, transport responses arising from $\mathbf{B}_{\rm emer}$ are expected to correlate with the creation, annihilation, or chirality reversal of skyrmion textures, unlike ordinary external magnetic-field-induced responses. Furthermore}, 
in our framework, although $\mathbf{B}_\mathrm{emer}$ is identical for both Weyl nodes,
its projection onto the Berry curvature
$\mathbf{B}_{\mathrm{emer}}\!\cdot\!\boldsymbol{\Omega}^{\chi}_\mathbf{k}$
automatically acquires opposite signs for opposite chiralities due to
$\boldsymbol{\Omega}^{\chi}_\mathbf{k}\propto \chi$~\cite{ahmad2021longitudinal}.
The resulting transport corrections, therefore, mimic axial-field-like
magnetotransport without invoking any explicit chiral gauge field.
This demonstrates that axial-type longitudinal and planar Hall
responses can arise purely from the coexistence of momentum-space
monopoles and real-space skyrmion topology. Our results thus establish a direct bridge between two distinct topological structures: the Weyl monopole in momentum space and the
skyrmion winding number in real space. The magnetotransport response
can be interpreted as a mixed topological coupling proportional to the product of these two topological charges, providing a unified description of anomaly-like transport phenomena driven by magnetic texture topology. 

\textcolor{black}{In the present work, the interplay between real-space and momentum-space topology is investigated within the semiclassical transport regime, where the dominant effect of the magnetic texture is captured through the emergent magnetic field $\mathbf{B}_{\rm emer}$. We note that the appearance of additive or multiplicative transport contributions does not necessarily imply the absence of interplay between the two topological structures. Indeed, recent phase-space Berry-curvature formulations treating real-space and momentum-space coordinates on equal footing have shown that the leading transport responses can remain effectively separable, while genuine mixed phase-space curvature corrections appear only at higher order~\cite{verma2022unified,addison2025anomalous,addison2023theory}. The present work, therefore, focuses on the leading semiclassical transport consequences of the coexistence of real-space magnetic topology and momentum-space Berry curvature in Weyl semimetals.}

\textcolor{black}{The semiclassical Boltzmann framework employed in the present work builds upon earlier transport formalisms developed for magnetic-field-driven transport in topological systems~\cite{knoll2020negative,sharma2017chiral,sharma2020sign,ahmad2021longitudinal,ahmad2023longitudinal,ahmad2024chiral,ahmad2024geometry,ahmad2025chiral,ahmad2024chiral,mandal2025chiral,mandal2025chiral,mandal2025disentangling,ghosh2024linear,mandal2026longitudinal,mandal2506disentangling,mandal2506distinguishing,mandal2026unconventional}. In this work, our focus is to investigation of Weyl-semimetal magnetotransport in the simultaneous presence of external and texture-induced emergent magnetic fields generated by topologically nontrivial magnetic textures. Specifically, we analyze how emergent electrodynamics associated with real-space magnetic topology modifies Berry-curvature- and chiral-anomaly-driven transport responses within the beyond-RTA framework.}

The remainder of the paper is organized as follows. In Sec.~\ref{Sec:Model}, we introduce the minimal model of a time-reversal symmetry–broken WSM and Sec.~\ref{sec:App_MBT} outline the semiclassical Boltzmann formalism incorporating Berry-curvature corrections and the emergent magnetic field arising from real-space topology. In Sec.~\ref{Sec.:Results and discussions}, we classify different regimes of sign reversal and present detailed results for the LMC and PHC in the presence of mixed real- and momentum-space topology. Finally, in Sec.~\ref{Sec.:Conclusion}, we summarize our main findings and discuss their physical implications.
\section{Low-Energy Model for Mixed Real- and Momentum-Space Topology}
\label{Sec:Model}
To investigate transport arising from the interplay of momentum-space and
real-space topology, we begin with the minimal low-energy Hamiltonian
describing quasiparticles near a pair of Weyl nodes.
In the absence of magnetic texture, the effective Hamiltonian reads
\cite{yan2017topological,lv2015experimental,armitage2018weyl,hasan2021weyl}:
\begin{align}
H_{\mathrm{WSM}}^\chi(\mathbf{k})
=
\chi \hbar v_\mathrm{F} \, \mathbf{k}\cdot\boldsymbol{\sigma},
\label{Hamiltonian}
\end{align}
where $\chi=\pm1$ denotes the chirality (topological charge) of the Weyl node,
$v_\mathrm{F}$ is the Fermi velocity, and $\boldsymbol{\sigma}$ represents the Pauli
matrices acting in pseudospin space.
The wave vector $\mathbf{k}$ is measured from the Weyl node.
For clarity and to isolate the essential mixed-topological contribution,
we restrict our analysis to a single pair of Weyl cones, noting that
additional pairs modify only the total response quantitatively
\cite{ahmad2021longitudinal,ahmad2024geometry}. Diagonalization of Eq.~\eqref{Hamiltonian} yields the linear dispersion
\begin{align}
\epsilon_{\mathbf{k}} = \pm \hbar v_\mathrm{F} |\mathbf{k}|.
\label{Dispersion}
\end{align}
The orbital magnetic moment (OMM), arising from the self-rotation of the Bloch wave packet, plays an important role in the semiclassical dynamics. It is given by~\cite{xiao2010berry}:
\begin{align}
\mathbf{m}^{\chi}_\mathbf{k}&= -\frac{ie}{2\hbar} \mathrm{Im} 
\left\langle \nabla_{\mathbf{k}} u^{\chi}(\mathbf{k}) \left| 
\left[  \epsilon_\mathbf{k} - \hat{H}^{\chi}_\mathrm{WSM}(\mathbf{k}) \right] 
\right| \nabla_{\mathbf{k}} u^{\chi} (\mathbf{k}) \right\rangle \nonumber \\
&\equiv -{\chi e v_\mathrm{F}}\mathbf{k}/{2k^2},
\label{Eq:OMM_formula}
\end{align}
where $u^{\chi}(\mathbf{k})$ is the Bloch wave packet in the WSM. The coupling of the OMM to a total magnetic field $\mathbf{B}_{\mathrm{tot}}$ modifies the band dispersion according to $\epsilon^{\chi}_{\mathbf{k}}=\epsilon_{\mathbf{k}}-\mathbf{m}^{\chi}_{\mathbf{k}}\cdot\mathbf{B}_{\mathrm{tot}}$.
Consequently, the constant-energy contour at the Fermi energy $\epsilon_\mathrm{F}$ is obtained as~\cite{ahmad2021longitudinal,sharma2020sign,varma2026chiral}:
\begin{align}
k^{\chi}_\mathrm{F} = \frac{\epsilon_\mathrm{F}+\sqrt{\epsilon^{2}_\mathrm{F} -  n \chi e v_\mathrm{F} B_{\mathrm{tot}} \beta_{\theta \phi}}}{n},
\end{align}
where $n = 2 \hbar v_\mathrm{F}$. By using $\theta$ and $\phi$ as polar and azimuthal angles of the Bloch eigenvectors, respectively (see below), the geometric factor $\beta_{\theta \phi}$ is given by  
\begin{align}
\beta_{\theta \phi} &= \sin{\theta} \cos{\phi} ~(\cos{\gamma} + \cos{\gamma}_{\mathrm{emer}}) \nonumber 
\\ &+ \cos{\theta} ~(\sin{\gamma} + \sin{\gamma}_{\mathrm{emer}}),
\end{align}
where the direction of the magnetic fields is assumed in the $xz$ plane, measured with respect to the $\hat{x}$-axis, i.e., $\mathbf{B} = B (\cos\gamma, 0, \sin\gamma)$ and $\mathbf{B}_{\mathrm{emer}} = B_{\mathrm{emer}} (\cos\gamma_{\mathrm{emer}}, 0, \sin\gamma_{\mathrm{emer}})$. The nontrivial topology of the Weyl node is encoded in its Bloch eigenvectors,
\begin{align}
\ket{u^{+}}^{T}&=[e^{-i\phi}\cos(\theta /2),~~\sin(\theta/2)], \\
\ket{u^{-}}^{T}&=[-e^{-i\phi}\sin(\theta /2),~~\cos(\theta/2)],
\end{align}
which generate a singular Berry curvature in momentum space,
\begin{align}
\boldsymbol{\Omega}^{\chi}_{\mathbf{k}}
= i \nabla_{\mathbf{k}} \times \langle u^{\chi}(\mathbf{k}) | \nabla_{\mathbf{k}} | u^{\chi}(\mathbf{k}) \rangle \equiv
-\chi \frac{\mathbf{k}}{2k^{3}},
\label{Berry_curvature}
\end{align}
acting as a monopole of Berry flux located at the Weyl node
\cite{xiao2010berry,son2012berry}.
This momentum-space Berry curvature underlies anomalous transport
responses such as the anomalous Hall effect and chiral-anomaly-induced
magnetotransport~\cite{ahmad2025longitudinal,ahmad2021longitudinal,ahmad2024geometry,ahmad2025chiral,varma2026chiral,nandy2017chiral,nandy2021chiral,zhang2016linear,yang2015chirality,sharma2020sign,PhysRevB.104.205124,das2019berry,das2020chiral,das2022nonlinear,sharma2019transverse,sharma2016nernst,sharma2017chiral,zeng2021nonlinear,sharma2017nernst,nandy2019planar,das2020thermal,das2020chiral,das2022nonlinear,das2023chiral,das2020chiral,mandal2022chiral,mandal2020thermopower,mandal2021tunneling,ghosh2024linear}. 

When the WSM hosts a slowly varying magnetic texture
$\mathbf{n}(\mathbf{r})$, such as a skyrmion lattice,
electrons traversing the texture acquire an additional real-space Berry phase.
In the adiabatic regime where the magnetic texture varies on a length scale
much larger than the electronic wavelength,
the effect of the texture can be captured through an emergent magnetic field
\cite{nagaosa2013topological,schulz2012emergent}, $\mathbf{B}_{\mathrm{emer}} = \frac{\hbar}{2e} \mathbf{n}\cdot (\partial_x \mathbf{n}\times\partial_y \mathbf{n}) ~\hat{e}$, which is directly proportional to the skyrmion density.
Unlike the external magnetic field, $\mathbf{B}_{\mathrm{emer}}$ originates
purely from real-space topology and exists even in the absence of
applied fields. In systems where Weyl band topology coexists with skyrmion textures, as realized in magnetic kagome materials and related chiral magnets~\cite{le2026ultra,dominici2023coupled,he2024enhanced,raju2024anisotropic}, charge carriers therefore experience \emph{two independent sources of
Berry curvature}: (i) Momentum-space Berry curvature $\boldsymbol{\Omega}^{\chi}_{\mathbf{k}}$
associated with Weyl nodes, and (ii) Real-space Berry curvature encoded in $\mathbf{B}_{\mathrm{emer}}$. Within the semiclassical description,
both fields enter the equations of motion through the modified
phase-space factor
\begin{align}
\mathcal{D}^{\chi}
=\left[1+\frac{e}{\hbar} (\mathbf{B}+\mathbf{B}_{\mathrm{emer}}) \cdot \boldsymbol{\Omega}^{\chi}_{\mathbf{k}} \right]^{-1}.
\label{phase_space}
\end{align}
The crucial new ingredient is the mixed scalar product
\(
\mathbf{B}_{\mathrm{emer}}\cdot
\boldsymbol{\Omega}^{\chi}_{\mathbf{k}},
\)
which vanishes unless both real-space and momentum-space topology are
simultaneously present. This term modifies the density of states and transport coefficients,
giving rise to additional contributions to LMC and PHC that are absent in
ordinary metals and in systems hosting only one type of topology. Thus, the effective low-energy model of Eq.~\eqref{Hamiltonian},
supplemented by the emergent field ${\mathbf{B}_\mathrm{emer}}$,
provides the minimal theoretical framework required to describe
\emph{mixed real–momentum space topological transport}. 

We assume that the length scale of the skyrmion textures is much smaller than the length of the system, so that the charge carrier experiences an average, uniform ${\mathbf{B}_\mathrm{emer}}$. This leads to the momentum-dependent but position-independent phase-space factor $\mathcal{D}^{\chi}_\mathbf{k}$. The magnitude of the emergent magnetic field generated by a skyrmion
texture is determined by the skyrmion density through
$B_{\mathrm{emer}} = n_{\mathrm{sk}}\Phi_0$, where $\Phi_0=h/e$ is the
flux quantum. For typical skyrmion sizes of $\sim 50$--$100\,\mathrm{nm}$
observed in chiral magnets, the corresponding emergent magnetic field
lies in the range $10^{-2}$--$1\,\mathrm{T}$. Therefore, the parameter
range $B_{\mathrm{emer}}=0$--$0.25\,\mathrm{T}$ used in our calculations
falls well within experimentally realistic values \cite{tokura2020magnetic}.
\section{Maxwell Boltzmann Transport formalism for electron transport in Skyrmion} 
\label{sec:App_MBT}
Using the quasiclassical Boltzmann transport theory, we study transport in WSMs in the limit of weak electric and magnetic fields. Since the quasiclassical Boltzmann theory is valid when the chemical potential $\mu$ is away from the nodal point such that $\mu^2 \gg \hbar {v_\mathrm{F}}^2 e |\mathbf{B}_{\mathrm{tot}}|$, we assume that $\mu$ lies in the conduction band without loss of generality. The phenomenological Boltzmann equation for the non-equilibrium distribution function $f^\chi_\mathbf{k}$ can be expressed as~\cite{kim2014boltzmann,knoll2020negative,imran2018berry,zyuzin2017magnetotransport,das2019linear,das2019berry} 
\begin{align}
\left(\frac{\partial}{\partial t} + \dot{\mathbf{r}}^\chi\cdot \nabla_\mathbf{r}+\dot{\mathbf{k}}^\chi\cdot \nabla_\mathbf{k}\right)f^\chi_\mathbf{k} = \mathcal{I}_{\mathrm{coll}}[f^\chi_\mathbf{k}],
\label{Eq_boltz1_mod}
\end{align}
where the collision term $\mathcal{I}_{\mathrm{coll}}$ incorporates impurity scattering. In the presence of an external magnetic field $\mathbf{B}$, a real-space emergent magnetic field $\mathbf{B}_{\mathrm{emer}}$, and an electric field $\mathbf{E}$, the semiclassical equations of motion become~\cite{sundaram1999wave,son2012berry,knoll2020negative,duval2006comment,xiao2010berry}
\begin{align}
\dot{\mathbf{r}}^\chi &= \mathcal{D}^\chi_\mathbf{k} 
\left[
\mathbf{v}^\chi_\mathbf{k}
+ \frac{e}{\hbar}\mathbf{E}\times \boldsymbol{\Omega}^\chi_\mathbf{k}
+ \frac{e}{\hbar}(\mathbf{v}^\chi_\mathbf{k}\cdot  \boldsymbol{\Omega}^\chi_\mathbf{k})\mathbf{B}_{\mathrm{tot}}
\right],
\nonumber\\
\dot{\mathbf{k}}^\chi &= -\frac{e}{\hbar} \mathcal{D}^\chi_\mathbf{k} 
\left[
\mathbf{E} + \mathbf{v}^\chi_\mathbf{k}\times \mathbf{B}_{\mathrm{tot}} 
+ \frac{e}{\hbar}(\mathbf{E}\cdot\mathbf{B}_{\mathrm{tot}})\boldsymbol{\Omega}^\chi_\mathbf{k}
\right],
\label{Eq.:EOM}
\end{align}
where $\mathbf{v}_\mathbf{k}^\chi$ is the band velocity calculated using $\epsilon^{\chi}_{\mathbf{k}}$. \textcolor{black}{The present study primarily focuses on the transport contributions arising from the chiral anomaly, while the conventional Lorentz-force contribution is neglected. More specifically, the Lorentz-force contribution enters the Boltzmann equation through the terms  $\propto[(\mathbf{v}^{\chi}_{\mathbf{k}}\times \mathbf{B}_{\mathrm{tot}})
\cdot \nabla_{\mathbf{k}} g^{\chi}_{\mathbf{k}}]$, which produces corrections proportional to the cyclotron parameter $\omega_c\tau$ ~\cite{bruus2004many,ziman1979principles,roy2016universal,ahmad2026semiclassical}:
\begin{align}
\omega_c\tau = \frac{eB}{m}\tau\equiv  \frac{e v_F^2 B}{\epsilon_F}\tau \ll 1,
\end{align}
where $\tau$ denotes the relevant transport relaxation time. In the weak-field metallic regime $\omega_c\tau\ll1$, these contributions generate higher-order angular distortions of the distribution function that remain subleading compared to the anomaly-induced contribution $\propto (\mathbf{E}\cdot\mathbf{B}_{\mathrm{tot}})\boldsymbol{\Omega}^{\chi}_{\mathbf{k}}$. Consequently, the dominant field-dependent contribution to the magnetotransport considered in the present work originates from the chiral anomaly terms. Similar approximations have been employed extensively in previous studies of anomaly-driven magnetotransport within the semiclassical regime~\cite{knoll2020negative,das2019linear,das2020thermal,ahmad2021longitudinal,sharma2016nernst,PhysRevB.107.115161,nandy2021chiral,li2021nonlinear,zeng2022chiral,ahmad2026semiclassical,kharzeev2014chiral,kumar2022chiral,johansson2019chiral,nandy2017chiral,johansson2019chiral,ahmad2026chiral}.} 

The collision integral includes both intra-node and inter-node scattering~\cite{mahan20089,bruus2004many,ziman1979principles,knoll2020negative,ahmad2024geometry}:
\begin{align}
\mathcal{I}_{\mathrm{coll}}[f^\chi_\mathbf{k}] 
= \sum_{\chi'}\sum_{\mathbf{k}'} W^{\chi\chi'}_{\mathbf{k}\mathbf{k}'}
(f^{\chi'}_{\mathbf{k}'} - f^\chi_\mathbf{k})
\label{Collision_integral}
\end{align}
with scattering rate calculated using Fermi's golden rule~\cite{ziman1979principles,bruus2004many,mahan20089,abers2004quantum}, 
\begin{align}
W^{\chi\chi'}_{\mathbf{k}\mathbf{k}'}
= \frac{2\pi}{\hbar} \frac{n_\mathrm{i}}{ \mathcal{V}}
|\langle u^{\chi'}(\mathbf{k}')| U^{\chi\chi'}_{\mathbf{k}\mathbf{k'}}|u^\chi(\mathbf{k})\rangle|^2 
\delta(\epsilon^{\chi'}_{\mathbf{k}'}-\epsilon_\mathrm{F}),
\end{align}
where $n_\mathrm{i}$ is the concentration of impurities, $\mathcal{V}$ is the volume of the system, and \(U^{\chi\chi'}_{\mathbf{k}\mathbf{k}'}\) is the profile of the scattering potential. 
 Assuming non-magnetic point-like impurities with momentum-independent and chirality-dependent amplitudes, i.e. $U^{\chi \chi'}_{\mathbf{k k'}} = U^{\chi\chi'}$, we can express the square of the matrix elements as
$\mathcal{G}^{\chi\chi'} \equiv |\bra{u^{\chi'}(\mathbf{k'})}U^{\chi \chi'}_{\mathbf{k k'}}\ket{u^{\chi}(\mathbf{k})}|^2 = {|U^{\chi\chi'}|^2\mathcal{V}^2}[1+\chi\chi'(\cos{\theta}\cos{\theta'} + \sin{\theta}\sin{\theta'}\cos{\phi}\cos{\phi'} + \sin{\theta}\sin{\theta'}\sin{\phi}\sin{\phi'})]$. Here, it is worth highlighting that, in our formalism, the relative strength of the intra- and inter-node scattering channels can be tuned using the parameter defined as: 
\begin{align}
    \alpha = \frac{U^{\chi\chi'}}{U^{\chi\chi}}\ \ \ (\chi'\neq\chi).
    \label{eq:alpha-definition}
\end{align}
\textcolor{black}{For the weak field regime considered here, we can expand the non-equilibrium distribution function as: $f^\chi_\mathbf{k} = { f_0} + g^\chi_\mathbf{k}$ with the Fermi distribution function $f_0$ and first order field induced deviation $g^\chi_\mathbf{k} = e \left(-\frac{\partial { f_0}}{\partial \epsilon^\chi_\mathbf{k}}\right) \mathbf{E} \cdot \mathbf{\Lambda}^\chi_\mathbf{k}$~\cite{morimoto2016semiclassical,sodemann2015quantum,das2022nonlinear,mandal2022chiral,dagnino2025non}. Here, $\mathbf{\Lambda}^\chi_\mathbf{k}$ is unknown function to be evaluated. Fixing $\mathbf{E}=E\hat{z}$, the component of $\mathbf{\Lambda}^\chi_\mathbf{k}$ along $\mathbf{E}$ would be relevant and one can write, $g^\chi_\mathbf{k} = e \left(-\frac{\partial { f_0}}{\partial \epsilon^\chi_\mathbf{k}}\right) E\Lambda^\chi_{\mathbf{k},z}$, where $\Lambda^\chi_{\mathbf{k},z}$ represents the $z$-component of unknown function (ansatz)~\cite{morimoto2016semiclassical,sodemann2015quantum,das2022nonlinear,mandal2022chiral}. For the sake of complexicity, we will drop the irrelevant index, $z$ in the $\Lambda^\chi_{\mathbf{k},z}$.} 

Keeping terms linear in $\mathbf{E}$, the Boltzmann equation becomes~\cite{bruus2004many,ziman1979principles,sharma2020sign,knoll2020negative,son2013chiral,sodemann2015quantum}:
\begin{align}
&\mathcal{D}^\chi_\mathbf{k}  \left[ v^{\chi z}_{\mathbf{k}} + \frac{e( B~\sin \gamma + B_{\mathrm{emer}}\sin \gamma_{\mathrm{emer}})}{\hbar}  ( \boldsymbol{\Omega}^\chi_\mathbf{k}\cdot \mathbf{v}^\chi_\mathbf{k})) \right] \nonumber\\&= \sum_{\chi'}\sum_{\mathbf{k}'} W^{\chi'\chi}_{\mathbf{k}\mathbf{k}'} (\Lambda^{\chi'}_{\mathbf{k}'} - \Lambda^\chi_\mathbf{k}).
\label{Eq_boltz3_mod}
\end{align}
Here, we define the band scattering rate as
\begin{align}
\frac{1}{\tau^{\chi}_{\mathbf{k}}(\theta,\phi)}=\sum_{\chi'}\mathcal{V}\int\frac{d^3\mathbf{k'}}{(2\pi)^3}(\mathcal{D}^{\chi'}_{\mathbf{k}'})^{-1} W^{\chi \chi'}_{\mathbf{k k'}}.
\label{Tau_invers}
\end{align}
Using Eq.~(\ref{Tau_invers}), we can write Eq.~(\ref{Eq_boltz3_mod}) as
\begin{align}
&h^{\chi}_\mathbf{k}(\theta,\phi) + \frac{\Lambda^{\chi}_\mathbf{k}(\theta,\phi)}{\tau_\mathbf{k}^{\chi}(\theta,\phi)}\nonumber \\
&=\mathcal{V}\sum_{\chi'}\int { \frac{d^3\mathbf{k}'}{(2\pi)^3} (\mathcal{D}^{\chi'}_\mathbf{k'})^{-1}} W^{\chi\chi'}_{\mathbf{k}\mathbf{k}'} \Lambda^{\chi'}_\mathbf{k'}(\theta',\phi') .
\label{MB_in_term_Wkk'_1}
\end{align}
Here, $h^{\chi}_\mathbf{k}(\theta,\phi)=\mathcal{D}^{\chi}_{\mathbf{k}}[v^{\chi}_{z,\mathbf{k}}+\frac{eB\sin{\gamma} + B_{emer}\sin{\gamma_{emer}}}{\hbar}(\mathbf{\Omega}^{\chi}_{k}\cdot \mathbf{v}^{\chi}_{\mathbf{k}})]$ and is evaluated at the Fermi energy. In the low temperature limit, for a constant Fermi energy surface, Eq.~(\ref{Tau_invers}) and the right-hand side of Eq.~(\ref{MB_in_term_Wkk'_1}) are reduced to integration over $\theta'$ and $\phi'$,
\begin{align}
&\frac{1}{\tau^{\chi}_\mathbf{k}(\theta,\phi)} =  \mathcal{V}\sum_{\chi'}\Pi^{\chi\chi'} \nonumber \\ &\ \ \ \ \ \ \ \ \ \ \ \ \ \ \  \times\iint\frac{(k'^{\chi'}_{\mathrm{F}})^3\sin{\theta'}}{|{\mathbf{v}^{\chi'}_{\mathbf{k'}_\mathrm{F}}}\cdot \mathbf{k'}_\mathrm{F}^{\chi'}|}d\theta'd\phi' ~\mathcal{G}^{\chi\chi'}(D^{\chi'}_{\mathbf{k'}})^{-1} \nonumber \\
&=\mathcal{V}\sum_{\chi'} \Pi^{\chi\chi'}\iint f^{\chi'}(\theta',\phi') ~\mathcal{G}^{\chi \chi'} \Lambda^{\chi'}_\mathbf{k}/\tau^{\chi}_\mathbf{k}~ d\theta' d\phi',
\label{Tau_inv_int_thet_phi}
\end{align}
{where} $\Pi^{\chi \chi'} = n_i / 4\pi^2 \hbar^2$ and $f^{\chi} (\theta,\phi)=\frac{(k_\mathrm{F}^{\chi})^3}{|{\mathbf{v}^\chi_{\mathbf{k}_\mathrm{F}}}\cdot \mathbf{k_\mathrm{F}^{\chi}}|} \sin\theta~ (\mathcal{D}^\chi_{\mathbf{k}})^{-1} ~\tau^{\chi}_\mathbf{k}(\theta,\phi)$. Using ansatz $\Lambda^{1,\chi}_{\mathbf{k}}=(d^{\chi}-h^{\chi}_\mathbf{k'} + a^{\chi}\cos{\theta} +b^{\chi}\sin{\theta}\cos{\phi}+c^{\chi}\sin{\theta}\sin{\phi})~\tau^{\chi}_\mathbf{k}(\theta,\phi)$, Eq.~(\ref{Tau_inv_int_thet_phi}) can be written in the following form:
\begin{multline}
d^{\chi}+a^{\chi}\cos{\theta}+b^{\chi}\sin{\theta}\cos{\phi}+c^{\chi} \sin{\theta}\sin{\phi}\\
=\sum_{\chi'}\Pi^{\chi\chi'}\iint f^{\chi'}(\theta',\phi') ~\mathcal{G}^{\chi\chi'} d\theta'd\phi'\\\times(d^{\chi'}-h^{\chi'}_\mathbf{k'}+a^{\chi'}\cos{\theta'}+b^{\chi'}\sin{\theta'}\cos{\phi'}+c^{\chi'} \sin{\theta'}\sin{\phi'}).\\
\label{Boltzman_final}
\end{multline}
When the aforementioned equation is explicitly stated, it appears as seven simultaneous \textcolor{black}{independent} equations that must be solved for eight variables. {The conservation of particle numbers} provides an additional restriction:
\begin{align}
\sum\limits_{\chi}\sum\limits_{\mathbf{k}} g^{\chi}_\mathbf{k} = 0.
\label{Eq_sumgk}
\end{align} 
For eight unknown variables ($d^{\pm 1}, a^{\pm 1}, b^{\pm 1}, c^{\pm 1}$), Eqs.~(\ref{Boltzman_final}) and (\ref{Eq_sumgk}) are simultaneously solved with Eq.~(\ref{Tau_inv_int_thet_phi}). Due to the intricate structure of {the} equations, all two-dimensional integrals with respect to $\theta'$ and $\phi'$ are carried out numerically and the { solution of simultaneous equations} are obtained. 

Once the distribution function $f^{\chi}_{\mathbf{k}}$ is evaluated, the current density can be evaluated as:
\begin{align}
    \mathbf{J} =-e\sum_{\chi,\mathbf{k}} f^{\chi}_{\mathbf{k}} ~\dot{\mathbf{r}}^{\chi}.
    \label{Eq:J_formula}
\end{align}
The generalized velocity $\dot{\mathbf{r}}^\chi_{\mathbf{k}}$ can be written as $\dot{\mathbf{r}}^\chi =\mathbf{v}^{\chi}_0 + \mathbf{v}^{\chi}_1$, where $\mathbf{v^{\chi}_{0}}$ and $\mathbf{v^{\chi}_{1}}$ correspond to the velocities without and with the electric field, respectively. { The} Explicit expressions are
\begin{align}
&\mathbf{v}^{\chi}_0 = \left(\frac{e\mathcal{D}^\chi_\mathbf{k}}{\hbar}\right) \left( \mathbf{v}_\mathbf{k}^\chi \cdot \boldsymbol{\Omega}^\chi_\mathbf{k}~ \mathbf{B}_{\mathrm{tot}} + \frac{\hbar}{e}\mathbf{v}_\mathbf{k}^\chi \right),\\\nonumber
&\mathbf{v}^{\chi}_1 = \left(\frac{e\mathcal{D}^\chi_\mathbf{k}}{\hbar}\right) \left( \mathbf{E} \times \boldsymbol{\Omega}^\chi_\mathbf{k} \right).
\end{align}
Now, from Eq.~(\ref{Eq:J_formula}), the generalized current { response} is given by
\begin{align}
   \mathbf{J} &=-e\sum_{\chi,\mathbf{k}}\left(\mathbf{v}^{\chi}_0f_0^\chi +\mathbf{v}^{\chi}_1~f_{0}^\chi 
   + \mathbf{v}^{\chi}_0~g^\chi_{\mathbf{k}}\right).
   \label{Eq:current upto E_sqr}
\end{align}
In the next sections, we evaluate LMC, $\sigma_{zz}$, and PHC, $\sigma_{xz}$, defined as $J_z=\sigma_{zz}E$ and $J_x=\sigma_{xz}E$, respectively.
{ Unless otherwise stated, we use the following parameter values: Fermi velocity $v_\mathrm{F} = 10^{6}\,\mathrm{m/s}$ and Fermi energy $\epsilon_\mathrm{F} = 0.05\,\mathrm{eV}$.} 

\begin{figure}
    \centering   \includegraphics[width=0.95\columnwidth]{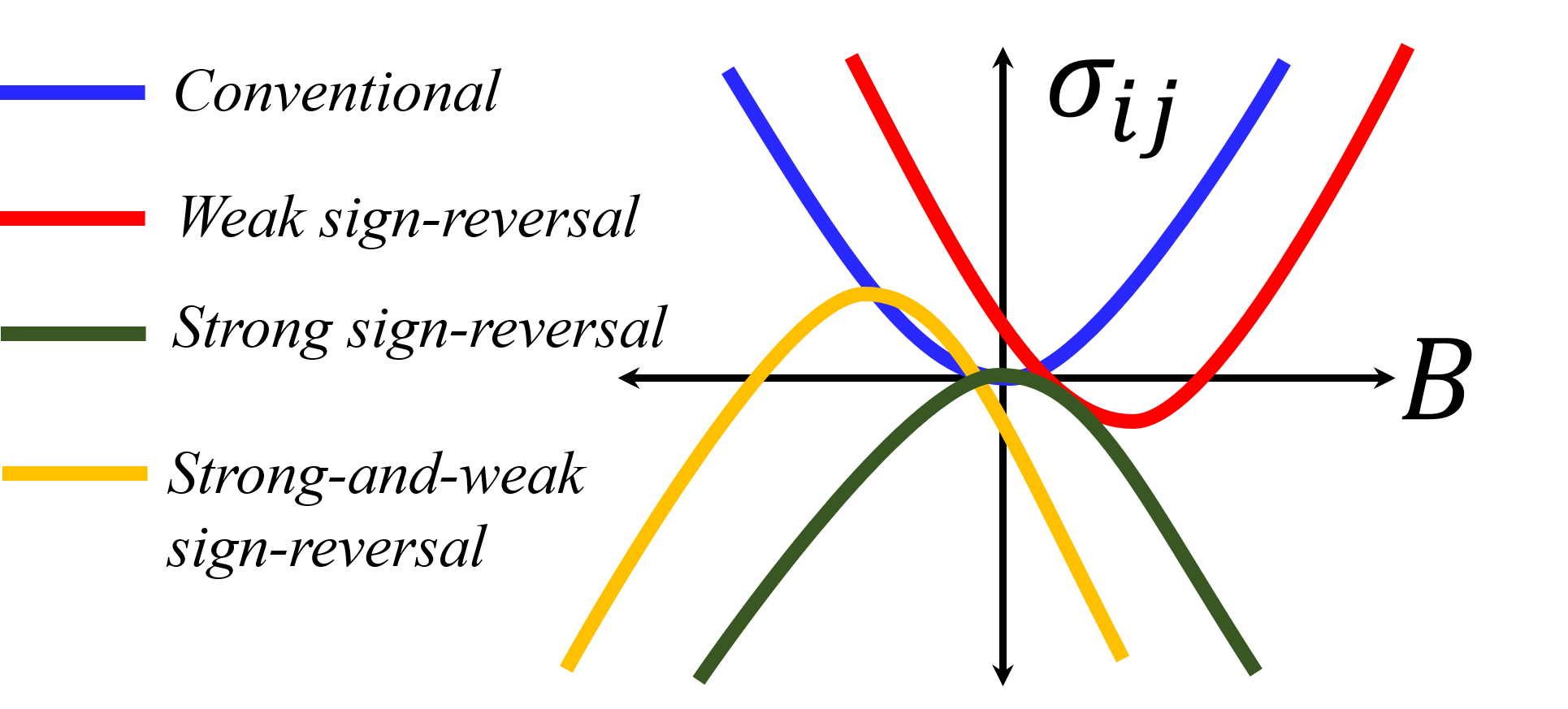}
    \caption{A schematic illustration of the characteristic regimes of magnetoconductivity $\sigma_{ij}$ in WSMs in the presence of both external magnetic field $B$ and emergent field $B_\mathrm{emer}$. The figure highlights \emph{weak sign reversal} (shifted parabola), \emph{strong sign reversal} (curvature inversion), and the coexisting \emph{strong-and-weak sign-reversal} regime, in contrast to the conventional quadratic-in-$B$ response. In the present mixed-topology setting, these regimes arise from the interplay between intervalley scattering and the real-space Berry-curvature contribution encoded in $B_{\mathrm{emer}}$ ~\cite{ahmad2023longitudinal,varma2024magnetotransport,varma2026chiral}.}  \label{fig:signreverseschematic}
\end{figure}
\section{Results and discussions}
\label{Sec.:Results and discussions}
Before presenting the detailed numerical results, we first clarify the
classification of different types of sign reversal of LMC in the present mixed-topology setting.
The sign of LMC in WSMs has been extensively investigated in
earlier works~\cite{knoll2020negative,sharma2020sign,sharma2023decoupling}. 
In an untilted WSM, the chiral anomaly typically
produces a positive LMC under weak intervalley scattering; however,
when intervalley scattering becomes sufficiently strong, the LMC
undergoes a sign reversal.
This mechanism originates from the competition between anomaly-driven
charge pumping and relaxation processes.
\begin{figure*}
    \centering   \includegraphics[width=2\columnwidth]{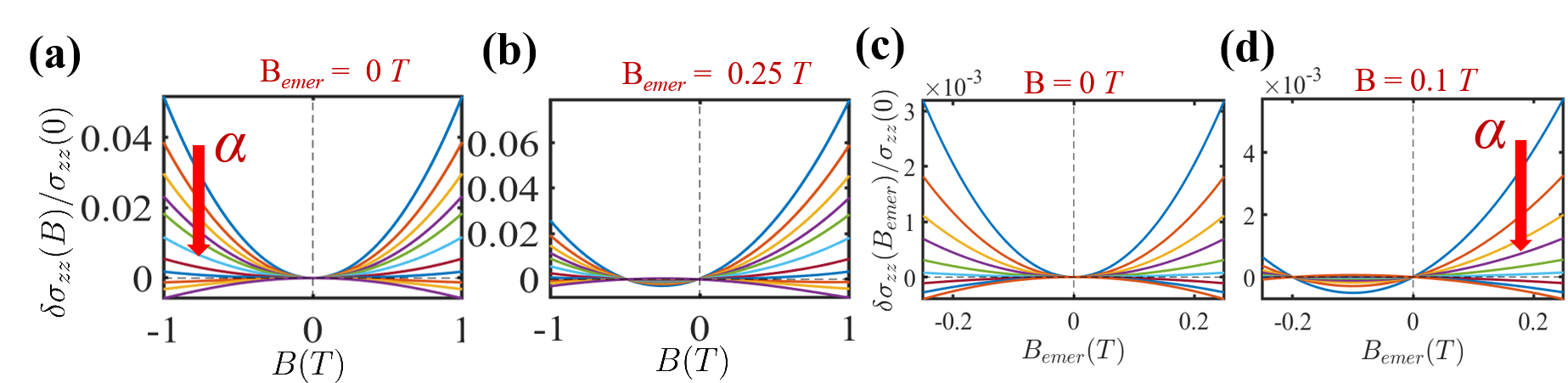}
    \caption{Normalized LMC for a minimal model of a time-reversal symmetry broken, untilted WSM in the presence of an emergent magnetic field $\mathbf{B}_\mathrm{emer}$ arising from real-space topology. The angle of $\mathbf{B}_\mathrm{emer}$ and an external magnetic field $\mathbf{B}$ is set to be $\gamma_\mathrm{emer}=\gamma=\pi/2$. In all the panels, $\alpha$ varies from 0.20 to 1.50 along the red arrow. (a) $B_\mathrm{emer}=0$~T and changing $B$. At $\alpha_c\simeq$~0.80, the anomaly-driven sign change of LMC occurs~\cite{ahmad2025longitudinal}. (b) $B_\mathrm{emer}=0.25$~T and changing $B$. For $\alpha<\alpha_c$ (${ \alpha}>\alpha_c)$, weak sign-reversal (strong-and-weak sign-reversal) behaviors emerge. (c) $B=0$~T and changing $B_\mathrm{emer}$. The sign change occurs at ${ \alpha_c}$. (d) $B=0.1$~T and changing $B_\mathrm{emer}$.} 
    \label{fig:LMC_vs_B_and_Be_vary_alp_vary}
\end{figure*}
\begin{figure}
    \centering
    \includegraphics[width = .98\columnwidth]{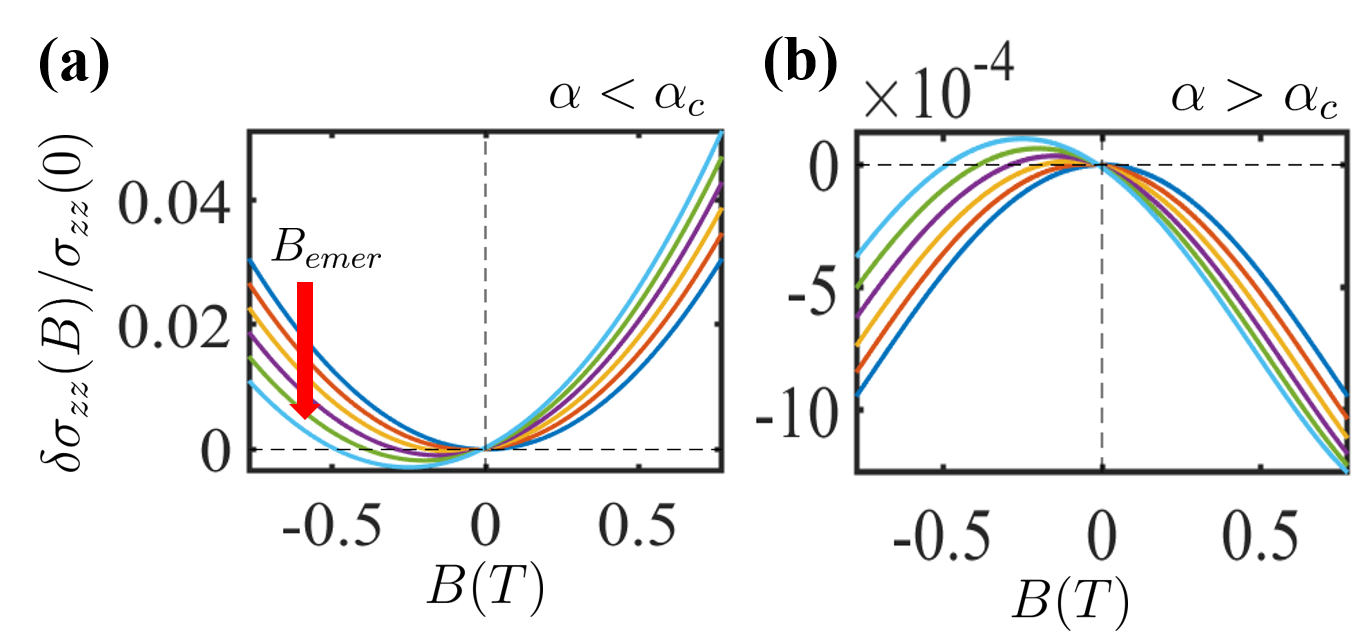}
    \caption{Normalized LMC as a function of the external magnetic field $\mathbf{B}$ with $\gamma= \pi/2$ for a minimal model of an untilted time-reversal symmetry broken WSM. (a) Weak sing-reversal regime ($\alpha=0.2 < \alpha_c { \simeq}$~0.80)  and (b) strong-and-weak sing-reversal regime ($\alpha=0.90 > \alpha_c$). The progression from the black to the green curves (indicated by the red arrow) corresponds to increasing the emergent magnetic field $B_\mathrm{emer}$ from 0 to 0.25~T with $\gamma_\mathrm{emer}=\pi/2$.}
    \label{fig:LMC_vs_B_Be_vary}
\end{figure}

In the present work, an additional ingredient enters the problem to enrich the dynamics of the electrons: the emergent magnetic field $\mathbf{B}_{\mathrm{emer}}$, which represents the
coarse-grained real-space Berry curvature generated by a skyrmion
texture. Although $\mathbf{B}_{\mathrm{emer}}$ enters the semiclassical equations
of motion in the same manner as an external magnetic field $\mathbf{B}$ through
$\mathbf{B}_{\mathrm{tot}}=\mathbf{B}+\mathbf{B}_{\mathrm{emer}}$,
its { role} is fundamentally distinct. $\mathbf{B}_{\mathrm{emer}}$ encodes real-space
topology and couples to the Weyl-node Berry curvature $\boldsymbol{\Omega}^{\chi}_{\mathbf{k}}$ via the modified
phase-space factor $\mathcal{D}^{\chi}$.

To characterize these behaviors systematically, we adopt the general
quadratic fitting form of the magnetoconductivity tensor~\cite{ahmad2023longitudinal}
\begin{align}
\sigma_{ij}(B) = \sigma_{ij}^{(0)} + (B-B_0)^2 \sigma_{ij}^{(2)},
\label{Eq-sij-fit}
\end{align}
which provides a unified framework to describe (i) conventional
quadratic-in-$B$ magnetoconductivity, (ii) shifted parabolic behavior,
and (iii) complete inversion of the curvature{, as schematically illustrated in Fig.~\ref{fig:signreverseschematic}}. 
Within this parametrization, we define the following three regimes.
\textit{Weak-sign-reversal} is characterized by
(i) $B_0 \neq 0$, 
(ii) $\sigma_{ij}^{(0)} \neq \sigma_{ij}(B=0)$, and 
(iii) $\mathrm{sign}\,(\sigma_{ij}^{(2)})>0$.
In this case, the vertex of the parabola in $\sigma_{ij}(B)$ shifts
away from the origin, and $\sigma_{ij}(B)$ exhibits opposite signs
for either small positive or negative magnetic fields. Importantly,
the overall curvature remains positive.
In the present context, such behavior can arise from the additional
real-space topological contribution induced by a finite
$B_{\mathrm{emer}}$, which effectively shifts the balance between
anomaly-driven enhancement and scattering-induced suppression.
\textit{Strong-sign-reversal}, on the other hand, corresponds to
$\mathrm{sign}\,(\sigma_{ij}^{(2)})<0$ with $B_0=0$, indicating a complete inversion
of the parabolic curvature. This regime is typically driven by strong
intervalley scattering, which overwhelms the anomaly contribution and
reverses the overall magnetic-field response.
Finally, when both a curvature inversion
($\mathrm{sign}\,(\sigma_{ij}^{(2)})<0$) and a finite shift ($B_0 \neq 0$)
occur simultaneously, the system exhibits a \textit{strong-and-weak
sign-reversal} regime. In the mixed-topology scenario considered here,
this regime naturally emerges from the cooperative interplay between
momentum-space Berry curvature (Weyl monopoles) and real-space
topology (skyrmion-induced emergent field). We are now in the position to discuss the magnetoconductivity in the presence of real and momentum space topology.\\
\subsection{LMC in the presence of real- and momentum-space topology}
\label{Sec:LMC in the presence of real and momentum space topology}

We begin by recovering the well-established behavior of the LMC in the absence of real-space topology
($B_{\mathrm{emer}}=0$). Figure~\ref{fig:LMC_vs_B_and_Be_vary_alp_vary}(a) shows the normalized LMC defined as 
\begin{align}
\frac{\delta\sigma_{zz}(B)}{\sigma_{zz}(0)}=\frac{\sigma_{zz}(B)-\sigma_{zz}(0)}{\sigma_{zz}(0)}
\label{normalized-LMC}
\end{align}
for the angle $\gamma=\pi/2$. Increasing the intervalley scattering strength $\alpha$ drives a sign
reversal of the LMC across the critical value $\alpha_c$. For
$\alpha < \alpha_c$, the anomaly-induced positive LMC dominates,
whereas for $\alpha > \alpha_c$, strong intervalley relaxation
overwhelms the chiral pumping mechanism and inverts the curvature of the magnetoconductivity~\cite{knoll2020negative,sharma2020sign,ahmad2021longitudinal,ahmad2023longitudinal,ahmad2024geometry,ahmad2025chiral,ahmad2025longitudinal,varma2026strain,varma2024magnetotransport,varma2026chiral}. 

We now incorporate real-space topology through a finite emergent magnetic
field $\mathbf{B}_\mathrm{emer}$ with $\gamma_\mathrm{emer}=\pi/2$, representing the coarse-grained Berry curvature
generated by a skyrmion texture. The corresponding LMC is shown in
Fig.~\ref{fig:LMC_vs_B_and_Be_vary_alp_vary}(b). Even an infinitesimal
$B_{\mathrm{emer}}$ qualitatively enriches the magnetic-field dependence
of the LMC: the vertex of the parabola shifts away
from the origin while its overall curvature remains unchanged for fixed
$\alpha$. According to the classification discussed earlier, this behavior
corresponds to \emph{weak sign reversal}. The shift originates from terms
linear in $B_{\mathrm{emer}}$ that arise through the modified phase-space
factor $\mathcal{D}^{\chi}_\mathbf{k}$.
Physically, the emergent field modifies the effective magnetic environment
experienced by Weyl quasiparticles without altering the intrinsic anomaly
mechanism.

Importantly, for a fixed value of $B_{\mathrm{emer}}$, 
increasing $\alpha$ still leads to inversion of the parabola, i.e., a transition
to \emph{strong sign reversal}{, as shown in Fig.~\ref{fig:LMC_vs_B_and_Be_vary_alp_vary}(b)}. Thus, intervalley scattering controls the curvature of the LMC, while the emergent field controls the shift
of its extremum. The coexistence of these two effects naturally gives
rise to the \emph{strong-and-weak sign-reversal} regime. To gain further insight, we plot the normalized LMC for two fixed
values of $\alpha$ (one below and one above $\alpha_c$) at different
values of $B_{\mathrm{emer}}$ in Fig.~\ref{fig:LMC_vs_B_Be_vary}.
For $\alpha < \alpha_c$ [Fig.~\ref{fig:LMC_vs_B_Be_vary}(a)], increasing $B_{\mathrm{emer}}$ progressively shifts the parabola, producing \emph{weak sign reversal} without changing
its positive curvature. In contrast, for $\alpha > \alpha_c$ [Fig.~\ref{fig:LMC_vs_B_Be_vary}(b)],
the curvature is already inverted; the addition of
$B_{\mathrm{emer}}$ introduces an additional shift, leading to the
combined \emph{strong-and-weak sign-reversal} regime. These results
demonstrate that real-space topology and intervalley scattering
affect distinct geometrical features of the LMC curve.

We next analyze the LMC as a function of $B_{\mathrm{emer}}$ itself,
both in the absence and presence of an external magnetic field,
as shown in Figs.~\ref{fig:LMC_vs_B_and_Be_vary_alp_vary}(c) and \ref{fig:LMC_vs_B_and_Be_vary_alp_vary}(d), respectively.
In the absence of { an} external field, the LMC as a function of
$B_{\mathrm{emer}}$ exhibits { a} strong sign reversal, reflecting the
dominant role of the emergent field in controlling the curvature.
When an external magnetic field is present, the interplay between
$\mathbf{B}$ and $\mathbf{B}_{\mathrm{emer}}$ produces \emph{weak} or
\emph{strong-and-weak} sign reversal depending on the value of $\alpha$.
This confirms that $B_{\mathrm{emer}}$ acts as an independent
topological tuning parameter rather than merely re-normalizing the
external magnetic field.

It is important to emphasize that although $B_{\mathrm{emer}}$
enters the semiclassical equations of motion through the linear
combination $\mathbf{B}_{\mathrm{tot}}=\mathbf{B}+\mathbf{B}_{\mathrm{emer}}$
and has the same sign at both Weyl nodes, its transport effect is not
a trivial field renormalization. The key point is that
$B_{\mathrm{emer}}$ couples to the momentum-space Berry curvature
$\boldsymbol{\Omega}^{\chi}_{\mathbf{k}}$, which is opposite for
opposite chiralities. Consequently, the mixed term
$\mathbf{B}_{\mathrm{emer}}\!\cdot\!\boldsymbol{\Omega}^{\chi}_\mathbf{k}$
generates chirality-dependent corrections to the phase-space factor
and generalized velocity, thereby modifying the anomaly-driven LMC in a genuinely
topological manner. This mixed real- and momentum-space coupling
constitutes a central result of the present work.

\begin{figure}
    \centering
    \includegraphics[width = .98\columnwidth]{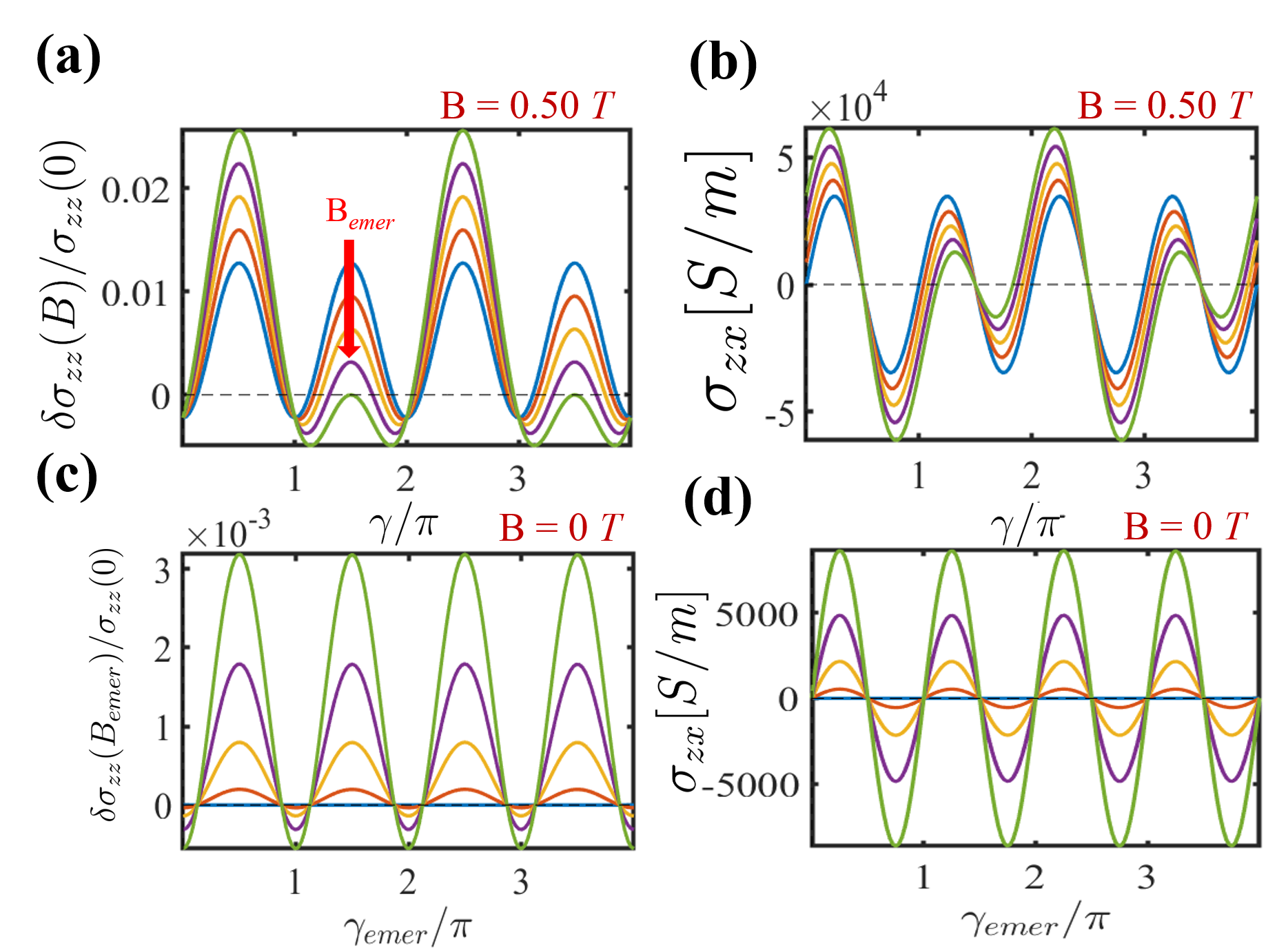}
    \caption{Normalized LMC and PHC for a minimal model of untilted WSM as a function of $\gamma$ and $\gamma_\mathrm{emer}$. (a) Normalized LMC at $B=0.5$~T. (b) PHC at $B=0.5$~T. (c) Normalized LMC for given $B_\mathrm{emer}$ at $B=0$~T. (d) PHC for given $B_\mathrm{emer}$ at $B=0$~T. In (a) and (b), from black to green lines, $B_{emer}$ is varied from $0$ to $0.25$~T with $\gamma_\mathrm{emer}=\pi/2$. In (c) and (d) $\gamma=\pi/2$}
    \label{fig:LMC_and_PHC_vs_gm_gmemr_piby2_Be_vary_alp_op2}
\end{figure}

Finally, we investigate the angular $\gamma$-dependence of the normalized LMC in the
presence of $B_{\mathrm{emer}}$ at $B=0.5$~T, shown in
Fig.~\ref{fig:LMC_and_PHC_vs_gm_gmemr_piby2_Be_vary_alp_op2}(a).
We observe that the emergent field introduces a pronounced asymmetry
in the angular response around $\gamma=\pi$. This asymmetry originates
from the fact that $B_{\mathrm{emer}}$ has the same sign at both Weyl
nodes and therefore breaks the effective inversion symmetry of the
magnetotransport response. The resulting angular dependence provides
an experimentally accessible signature of the mixed real-space and
momentum-space topology. To deepen our understanding of $B_\mathrm{emer}$ in LMC, we have plotted it as a function of $\gamma_\mathrm{emer}$ at different values of $B_\mathrm{emer}$ in Fig.~\ref{fig:LMC_and_PHC_vs_gm_gmemr_piby2_Be_vary_alp_op2}(c) at $B=0$. The angular anisotropy is clearly visible in this plot. Although the angular dependence remains intact for all values of $B_\mathrm{emer}$, the magnitude increases with increasing $B_{emer}$ and the positive portion increases more rapidly than negative one. As discussed earlier, due to the same sign of $B_\mathrm{emer}$ for both nodes, the fermi surface distorted along the direction of $B_\mathrm{emer}$ and this leads to an anisotropic response, and shows a maximum LMC for $\mathbf{B}_\mathrm{emer}||\hat{z}$.

\subsection{PHC in the presence of real- and momentum-space topology}
\label{Sec:PHC in the presence of real and momentum space topology}

Having discussed the LMC, we now turn to PHC. The PHC exhibits a characteristic
angular dependence proportional to $\sin(2\gamma)$, where $\gamma$ is
the angle of the applied magnetic field measured from the $x$-axis
\cite{nandy2017chiral}. In Fig.~\ref{fig:LMC_and_PHC_vs_gm_gmemr_piby2_Be_vary_alp_op2}(b), we plot this angular dependence for different values of $B_{\mathrm{emer}}$. In the presence of $B_{\mathrm{emer}}$, the angular profile becomes
asymmetric about $\gamma=\pi$. Although the functional dependence
remains similar for $\gamma \in (\pi,2\pi)$, the magnitude is reduced.
This asymmetry arises from the fixed orientation of $B_{\mathrm{emer}}$
along the $z$ direction. 

\begin{figure}
    \centering   \includegraphics[width=\columnwidth]{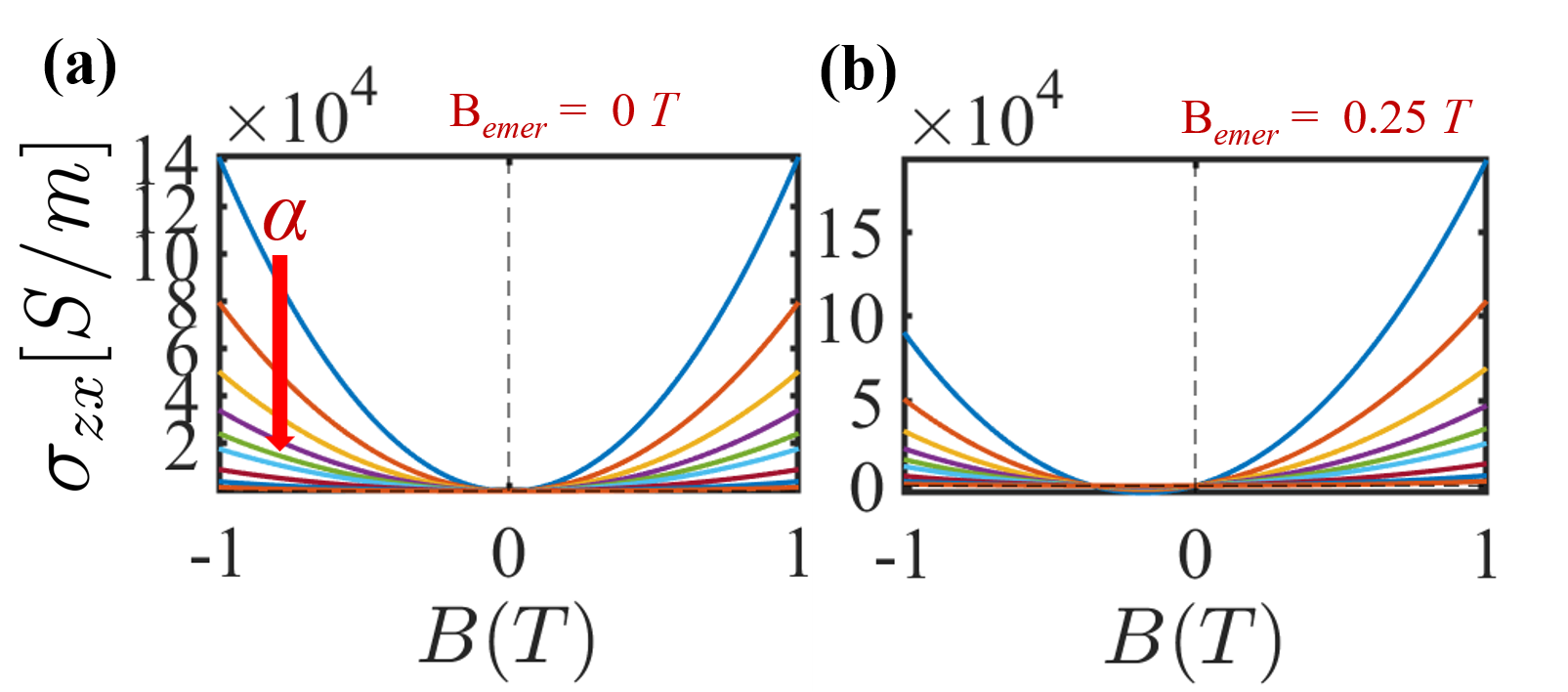}
    \caption{{PHC as a function of external magnetic field $B$ for a minimal model of an untilted time-reversal symmetry broken WSM. (a) The emergent magnetic field $B_\mathrm{emer}=0$~T. (b) $B_\mathrm{emer}=0.25$~T. The angle $\gamma = \pi/4$ and $\gamma_{emer} = \pi/2$. In (a) and (b), $\alpha$ is changed from 0.20 to 1.5 from the top to bottom lines.}}  \label{fig:PHC_vs_B_Be_vary_alp_vary}
\end{figure}

\begin{figure}
    \centering
    \includegraphics[width = .98\columnwidth]{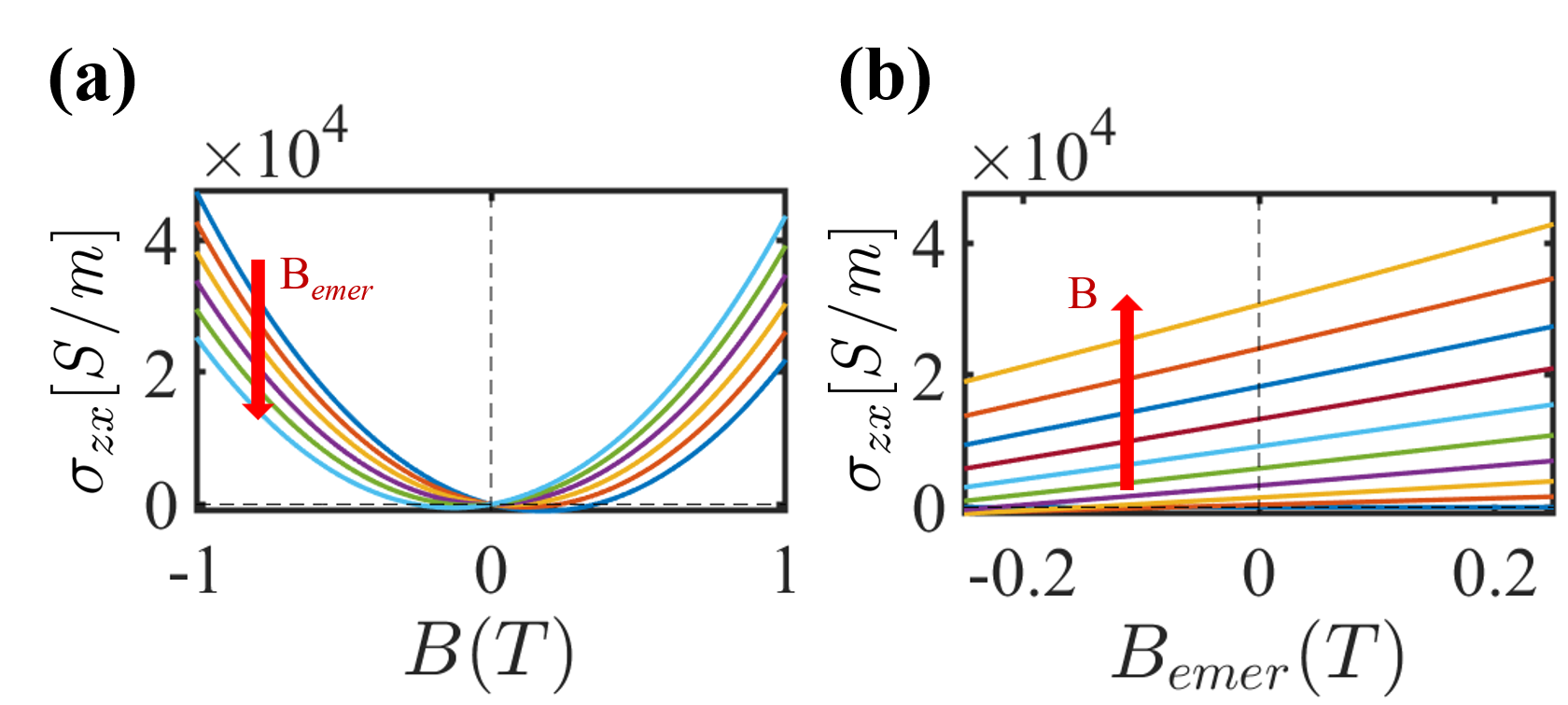}
    \caption{{PHC for a minimal model of an untilted time-reversal symmetry broken WSM. (a) PHC as a function of the external magnetic field $B$. Along the red arrow, the emergent magnetic field $B_\mathrm{emer}$ changes from $-$0.25~T to 0.25~T. (b) PHC as a function of $B_\mathrm{emer}$. Along the red arrow, $B$ changes from 0 to 1~T. In both, $\gamma = \pi/4$, $\gamma_\mathrm{emer} = \pi/2$, and $\alpha< \alpha_c$.}}
    \label{fig:PHC_vs_B_vary_Be_vary_alp_0p46}
\end{figure}

To isolate the role of real-space topology, we first evaluate the PHC
in the absence of $B_{\mathrm{emer}}$. Figure
\ref{fig:PHC_vs_B_Be_vary_alp_vary}(a) shows the planar Hall
conductivity $\sigma_{zx}$ for several values of the intervalley
scattering strength $\alpha$. In this case, the response is governed
purely by momentum-space topology. Increasing $\alpha$ suppresses the
magnitude of the PHC; however, unlike the LMC, no sign reversal is
observed. When $B_{\mathrm{emer}}$ is switched on
[Fig.~\ref{fig:PHC_vs_B_Be_vary_alp_vary}(b)], the minima of the PHC
parabola shift, but shows weak sign reversal. Fixing $\alpha$ and
varying $B_{\mathrm{emer}}$ leads to a qualitatively different
behavior, shown in
Fig.~\ref{fig:PHC_vs_B_vary_Be_vary_alp_0p46}(a). In this case, the
presence of real-space topology drives the system into a
weak-sign-reversal regime. In Fig.~\ref{fig:PHC_vs_B_vary_Be_vary_alp_0p46}(b), we plot the PHC as
a function of $B_{\mathrm{emer}}$ for different values of the external
magnetic field. For $B=0$, the PHC seems to vanish, indicating that an emergent field directed along $z$ alone does not generate a planar Hall response of the same magnitude as the external field does. Increasing the external magnetic field enhances the magnitude of the PHC within the studied parameter range. 

To gain further insight, we consider the PHC in the presence of $B_{\mathrm{emer}}$ alone while varying its direction within the $xz$-plane, as shown in Fig.~\ref{fig:LMC_and_PHC_vs_gm_gmemr_piby2_Be_vary_alp_op2}(d). Which shows PHC has very less magnitude, but still finite. It also confirm that $B_{\mathrm{emer}}$ can alone give a finite PHC. It is clear that $\sigma_{zx}\propto \sin 2\gamma_\mathrm{emer}$, which is intact for all finite values of $B_\mathrm{emer}$ taken. This has never been explored before! 

\section{Conclusions}
\label{Sec.:Conclusion}
\textcolor{black}{The present work extends beyond RTA WSM transport theory to the regime where emergent electrodynamics generated by real-space magnetic textures coexists with momentum-space topological transport.} We have investigated magnetotransport in a time-reversal symmetry-broken,
untilted WSM in the simultaneous presence of momentum-space
Berry curvature and real-space topology encoded through an emergent
magnetic field $\mathbf{B}_{\mathrm{emer}}$. Within a semiclassical Boltzmann
framework, we systematically analyzed the LMC and PHC in this mixed-topology setting. In the absence of real-space topology, we recover the known behavior:
the LMC undergoes a strong sign reversal
as intervalley scattering increases beyond a critical value, reflecting
the competition between anomaly-driven charge pumping and relaxation.
Upon introducing a finite $B_{\mathrm{emer}}$, the magnetoconductivity
acquires an additional shift of its parabolic profile, corresponding to
a weak-sign-reversal regime. While intervalley scattering controls the
curvature of the response, the emergent field controls the shift of its
extremum. Their combined action naturally generates a strong-and-weak
sign-reversal regime, demonstrating that real- and momentum-space
topology affect distinct geometrical features of the magnetoconductivity.

We further showed that the emergent field induces a pronounced asymmetry
in the angular dependence of both longitudinal and planar Hall
conductivities. For the planar Hall response, $\mathbf{B}_{\mathrm{emer}}$
modifies the magnitude and can drive weak-sign-reversal behavior,
while preserving the characteristic $\sin(2\gamma)$ dependence.
Notably, a finite PHC can arise solely from
{$\mathbf{B}_{\mathrm{emer}}$} when its direction is varied, confirming that
real-space topology alone can generate a measurable transverse response. Our results establish that $\mathbf{B}_{\mathrm{emer}}$ does not merely re-normalize the external magnetic field through
$\mathbf{B}_{\mathrm{tot}}=\mathbf{B}+{\mathbf{B}_{\mathrm{emer}}}$. Instead, through its coupling to momentum-space Berry curvature, it acts as an independent topological tuning parameter. The interplay between Weyl monopoles and skyrmion-induced emergent fields provides a direct transport signature of mixed real- and momentum-space topology. \textcolor{black}{Our results describe the leading semiclassical transport consequences arising from the coexistence of real-space magnetic topology and momentum-space Berry curvature, while higher-order mixed phase-space curvature effects remain an interesting direction for future work~\cite{verma2022unified,addison2025anomalous,addison2023theory}. Our results further suggest that transport signatures correlated with skyrmion chirality and texture evolution may provide a route to experimentally distinguish emergent-field-driven responses from conventional magnetic-field effects.}

Our results suggest clear experimental signatures that can be probed
using angle-dependent magnetotransport measurements in magnetic Weyl
semimetals hosting nontrivial spin textures. By rotating the external
magnetic field within the transport plane, one can measure the
LMC and PHC as
functions of both the field magnitude and orientation. The interplay
between Berry curvature in momentum space and the emergent magnetic
field generated by real-space topology is predicted to produce two
distinct features: (i) a finite shift of the magnetoconductivity
parabola away from $B=0$, and (ii) an angular asymmetry of the
magnetotransport response about $\gamma=\pi$. Since the emergent field
is directly related to the density of magnetic textures, these
transport signatures should evolve with the skyrmion density, which
can be independently characterized using magnetic imaging techniques
such as Lorentz transmission electron microscopy or magnetic force
microscopy. Observation of these features would provide direct
experimental evidence of mixed real- and momentum-space topology in
Weyl systems.

\section{Acknowledgments}
The authors are grateful to Amit Agarwal, Gautham Varma K, \textcolor{black}{Tanay Nag}, and Gargee Sharma for their insightful discussions and valuable scientific input. \textcolor{black}{A.A. acknowledges funding from the Core Research Grant by ANRF (Sanction No. CRG/2023/007003), Department of Science and Technology,
India}.  T.T. acknowledges financial support by the Japan Society for the Promotion of Science, KAKENHI (Grants No. 25H01248) from the Ministry of Education, Culture, Sports, Science, and Technology, Japan. The authors are very grateful to IIT Mandi for providing computing facilities.
\appendix
\section{Real and momentum space mixed term in the current}
\label{sec:Real and momentum space mixed term in the current}
We evaluate the current solely from the nonequilibrium correction, which is first order in $E$:
\begin{align}
\mathbf{J}
=
-e\sum_{\chi,\mathbf{k}}
\mathbf{v}_0^\chi\, g^{1,\chi}_{\mathbf{k}}.
\label{Eq:J_from_g1}
\end{align}
Using $\mathbf{v}_0^\chi = \mathcal{D}^\chi \left[ \mathbf{v}_\mathbf{k}^\chi + \frac{e}{\hbar} (\mathbf{v}_\mathbf{k}^\chi\!\cdot\!\boldsymbol{\Omega}^\chi) \mathbf{B}_{\mathrm{tot}} \right]$, the weak-field expansion gives $\mathbf{v}_0^\chi \simeq \mathbf{v}_\mathbf{k}^\chi + \frac{e}{\hbar} (\mathbf{v}_\mathbf{k}^\chi\!\cdot\!\boldsymbol{\Omega}^\chi)
\mathbf{B}_{\mathrm{tot}} - \frac{e}{\hbar} (\mathbf{B}_{\mathrm{tot}}\!\cdot\!\boldsymbol{\Omega}^\chi) \mathbf{v}_\mathbf{k}^\chi$. In constant relaxation-time approximation, $g^{1,\chi}_{\mathbf{k}}
= -e\tau \mathbf{E}\!\cdot\!\mathbf{v}_\mathbf{k}^\chi \left(-\frac{\partial f_0}{\partial \epsilon}\right)$. Keeping only terms linear in
$\mathbf{B}_{\mathrm{emer}}$, we obtain
\begin{align}
\mathbf{J}_{\mathrm{cross}}
&= \frac{e^3\tau}{\hbar}
\sum_{\chi,\mathbf{k}}\left(-\frac{\partial f_0}{\partial \epsilon}\right)
(\mathbf{E}\!\cdot\!\mathbf{v}_\mathbf{k}^\chi)
\nonumber\\
&\times\left[
(\mathbf{v}_\mathbf{k}^\chi\!\cdot\!\boldsymbol{\Omega}^\chi)
\mathbf{B}_{\mathrm{emer}}
-
(\mathbf{B}_{\mathrm{emer}}\!\cdot\!\boldsymbol{\Omega}^\chi)
\mathbf{v}_\mathbf{k}^\chi
\right].
\label{Eq:Jcross_g1}
\end{align}
For longitudinal transport, $\mathbf{E}=E_z\hat{z}$ and
$\mathbf{B}_{\mathrm{emer}}=B_{\mathrm{emer}}\hat{z}$.
Then
\begin{align}
J_z^{\mathrm{cross}}
&=
\frac{e^3\tau E_z B_{\mathrm{emer}}}{\hbar}
\nonumber\\
&\times\sum_{\chi,\mathbf{k}}
\left(-\frac{\partial f_0}{\partial \epsilon}\right) v_{k,z}^\chi \left[ (\mathbf{v}_\mathbf{k}^\chi\!\cdot\!\boldsymbol{\Omega}^\chi) - \Omega_z^\chi v_{k,z}^\chi \right].
\label{Eq:Jz_g1}
\end{align}
Thus the longitudinal conductivity acquires a term
\begin{align}
\sigma_{zz}^{\mathrm{cross}}
\propto
B\,B_{\mathrm{emer}},
\end{align}
which is linear in $B_{\mathrm{emer}}$ and linear in $B$ (through the
standard anomaly correction inside $g^1$ when solved self-consistently). Including the usual anomaly contribution $\propto B^2$,
the conductivity takes the general form
\begin{align}
\sigma_{zz}(B) = \sigma_{zz}^{(0)} + a_1 B^2 + b_1 B B_{\mathrm{emer}},
\label{Eq:sigma_structure}
\end{align}
where $a_1$ originates from the usual chiral anomaly and
$b_1$ arises from Eq.~(\ref{Eq:Jcross_g1}). Completing the square,
\begin{align}
\sigma_{zz}(B)
= \sigma_{zz}^{(0)} + a_1\left(B + \frac{b_1}{2a_1}B_{\mathrm{emer}}\right)^2 -
\frac{b_1^2}{4a_1}B_{\mathrm{emer}}^2,
\end{align}
Hence, the extremum shifts to $B_0 = -\frac{b_1}{2a_1} B_{\mathrm{emer}}$. For $\mathbf{B}=B(\cos\gamma \, \hat{x}+\sin\gamma \, \hat{z})$
and $\mathbf{B}_{\mathrm{emer}}=B_{\mathrm{emer}}\hat{z}$,
the mixed term in Eq.~(\ref{Eq:Jcross_g1}) produces
contributions proportional to $B B_{\mathrm{emer}}\sin\gamma$.
Although this term is formally invariant under $\gamma\to\gamma+\pi$
when the reversal of the magnetic field direction is taken into account,
the presence of a fixed $\mathbf{B}_{\mathrm{emer}}$ breaks the effective
$\gamma \to \gamma+\pi$ symmetry of the transport response.
As a result, $\sigma(\gamma) \neq \sigma(\gamma+\pi)$, leading to an
angular asymmetry originating from the real- and momentum-space
cross coupling. Physically, since $B_{\mathrm{emer}}$ has the same sign
at both Weyl nodes while $\boldsymbol{\Omega}^\chi$ changes sign with
chirality, the mixed contribution remains chirality dependent and
breaks the effective inversion symmetry of the angular magnetotransport response.
\color{black}
{\section{Phase-space Berry curvature analysis of mixed real- and momentum-space topology}
\label{app:phase_space}
In this appendix, we study the topologies by formulating the problem within the full phase-space Berry curvature within the full phase-space Berry-curvature framework. We show that the leading-order magnetotransport response is additive in the external magnetic field ${\bf B}$ and the Skyrmion-induced emergent magnetic field ${\bf B}_{\rm emer}$, while the genuine mixed phase-space Berry-curvature contribution is parametrically suppressed in the semiclassical regime.\\
For a Bloch electron moving in a slowly varying magnetic texture, the natural semiclassical coordinates are:
$\xi^\alpha=(r_x,r_y,r_z,k_x,k_y,k_z)$, and the dynamics is governed by the full phase-space Berry-curvature tensor
\begin{equation}
\Omega_{\alpha\beta}
=
\begin{pmatrix}
\Omega_{rr} & \Omega_{rk}\\
\Omega_{kr} & \Omega_{kk}
\end{pmatrix}.
\label{eq:Omega_full}
\end{equation}
Here $\Omega_{kk}$ denotes the momentum-space Berry curvature associated with the Weyl node,
$\Omega^{ij}_{kk} = \epsilon^{ijl}\Omega_l^\chi({\bf k})$, with $\bm{\Omega}^{\chi}({\bf k}) = -\chi \frac{{\bf k}} {2k^3}$,where $\chi=\pm 1$ labels the chirality. The real-space Berry curvature generated by a slowly varying Skyrmion texture is
\begin{equation}
\Omega_{rr}^{ij}
=
\frac12
\hat{\bf n}\cdot
\left(
\partial_{r_i}\hat{\bf n}
\times
\partial_{r_j}\hat{\bf n}
\right),\\
\label{eq:realspaceBC}
\end{equation}
which may be written in terms of the emergent magnetic field: $B_{{\rm emer},l}
=
\frac{\hbar}{e}
\frac12
\epsilon_{ijl}
\Omega_{rr}^{ij}$. Finally, the mixed phase-space Berry curvature is
\begin{equation}
\Omega_{kr}^{ij}
=
\frac12
\hat{\bf d}
\cdot
\left(
\partial_{k_i}\hat{\bf d}
\times
\partial_{r_j}\hat{\bf d}
\right),
\label{eq:mixed_BC}
\end{equation}
where ${\bf d}({\bf r},{\bf k})$ denotes the vector entering the local two-band Hamiltonian. For a slowly varying magnetic texture, momentum-space and real-space derivatives scale as $\partial_{k} \sim k_F^{-1}$, and $\qquad \partial_r \sim L_s^{-1}$, where $k_F$ is the Fermi wavevector and $L_s$ is the characteristic skyrmion length scale. Equation~(\ref{eq:mixed_BC}) therefore implies
\begin{equation}
\Omega_{kr}
\sim
\frac{1}{k_F L_s}.
\label{eq:mixed_scaling}
\end{equation}
The semiclassical regime requires $k_F^{-1} \gg L_s$, and hence
\begin{equation}
\Omega_{kr}
\ll
1.
\label{eq:smallmixed}
\end{equation}
Equation~(\ref{eq:smallmixed}) establishes that the mixed phase-space Berry curvature is parametrically small compared with the leading real-space and momentum-space Berry curvature contributions. A detailed analysis of the different regimes can be found in the literature~\cite{verma2022unified,addison2023theory,addison2025anomalous}.
\paragraph{Phase-space measure}
The invariant phase-space volume element is $dV = D(\xi)\, d^3r\,d^3k$, with $D(\xi) = \sqrt{\det \Omega}$. Expanding to the leading order in Berry curvatures gives
\begin{equation}
D(\xi)
= 1 + \frac{e}{\hbar} {\bf B}\cdot\bm{\Omega}^{\chi} +
\frac{e}{\hbar} {\bf B}_{\rm emer}\cdot\bm{\Omega}^{\chi}
+ {\rm Tr} \left[ \Omega_{kr}
\right] + \mathcal O(\Omega^2).
\label{eq:D_expand}
\end{equation}
The first correction arises from the ordinary magnetic field, the second from the real-space Berry curvature through the emergent field, while the third represents the genuine mixed phase-space contribution. Using Eq.~(\ref{eq:mixed_scaling}), the latter scales as ${\rm Tr}[\Omega_{kr}]
\sim \frac{1}{k_F L_s}$, and is therefore subleading in the semiclassical limit.
\paragraph{Conductivity}
The longitudinal conductivity can be written schematically as
\begin{equation}
\sigma_{ij} \sim
\int
[d{\bf k}]
\,D(\xi)\,
v_i
\Lambda_j ,
\label{eq:sigma_general}
\end{equation}
where $\Lambda_j$ denotes the solution of the Boltzmann equation. Substituting Eq.~(\ref{eq:D_expand}) into Eq.~(\ref{eq:sigma_general}) yields
\begin{equation}
\sigma_{ij}
=
\sigma_{ij}^{(0)}
+
\sigma_{ij}^{(B)}
+
\sigma_{ij}^{({\rm emer})}
+
\delta\sigma_{ij}^{(kr)}
+
\cdots ,
\label{eq:sigma_decomp}
\end{equation}
with $\sigma_{ij}^{(B)} \propto
{\bf B}\cdot\bm{\Omega}^{\chi},
\sigma_{ij}^{({\rm emer})} \propto
{\bf B}_{\rm emer}\cdot\bm{\Omega}^{\chi}$, and
$\delta\sigma_{ij}^{(kr)} \propto
\Omega_{kr}$. Using Eq.~(\ref{eq:mixed_scaling}), one finds $\delta\sigma_{ij}^{(kr)}
= \mathcal O
\left(
\frac{1}{k_F L_s}
\right)$, which is parametrically smaller than the leading contributions. Therefore, to leading order,
\begin{equation}
\sigma_{ij} = \sigma_{ij}^{(0)} + \sigma_{ij}^{(B)} + \sigma_{ij}^{({\rm emer})}
\label{eq:additive_result}
\end{equation}
and the conductivity is additive in the external and emergent magnetic fields.

\paragraph{Application to anomaly-induced longitudinal magnetoconductivity}

For a Weyl semimetal, the anomaly-induced conductivity depends on the total effective magnetic field ${\bf B}_{\rm tot} = {\bf B} + {\bf B}_{\rm emer}$. The leading longitudinal conductivity therefore, takes the form
\begin{equation}
\sigma_{zz}
=
\sigma_0
+
C
\left(
{\bf B}
+
{\bf B}_{\rm emer}
\right)^2 ,
\label{eq:lmc_total}
\end{equation}
where $C$ is determined by the band structure and intervalley scattering. Expanding Eq.~(\ref{eq:lmc_total}) gives
\begin{equation}
\sigma_{zz}
=
\sigma_0
+
C B^2
+
2C\,{\bf B}\cdot{\bf B}_{\rm emer}
+
C B_{\rm emer}^2.
\label{eq:lmc_expand}
\end{equation}
The cross term proportional to ${\bf B}\cdot{\bf B}_{\rm emer}$ arises solely from the square of the total effective magnetic field and survives even when $\Omega_{kr}=0$. Consequently, the term proportional to ${\bf B}\cdot{\bf B}_{\rm emer}$ does not represent a genuine mixed phase-space Berry-curvature effect. The true mixed contribution is instead contained in $\delta\sigma_{ij}^{(kr)}$, which is suppressed by the small parameter $(k_F L_s)^{-1}$. We therefore conclude that within the semiclassical regime $k_F^{-1}
\ll
L_s$, the leading magnetotransport response is governed by the additive contributions of momentum-space topology and real-space topology, while genuine mixed phase-space Berry-curvature effects enter only as higher-order corrections.}
\color{black}
\bibliography{biblio.bib}

\begin{thebibliography}{93}%
\makeatletter
\providecommand \@ifxundefined [1]{%
 \@ifx{#1\undefined}
}%
\providecommand \@ifnum [1]{%
 \ifnum #1\expandafter \@firstoftwo
 \else \expandafter \@secondoftwo
 \fi
}%
\providecommand \@ifx [1]{%
 \ifx #1\expandafter \@firstoftwo
 \else \expandafter \@secondoftwo
 \fi
}%
\providecommand \natexlab [1]{#1}%
\providecommand \enquote  [1]{``#1''}%
\providecommand \bibnamefont  [1]{#1}%
\providecommand \bibfnamefont [1]{#1}%
\providecommand \citenamefont [1]{#1}%
\providecommand \href@noop [0]{\@secondoftwo}%
\providecommand \href [0]{\begingroup \@sanitize@url \@href}%
\providecommand \@href[1]{\@@startlink{#1}\@@href}%
\providecommand \@@href[1]{\endgroup#1\@@endlink}%
\providecommand \@sanitize@url [0]{\catcode `\\12\catcode `\$12\catcode `\&12\catcode `\#12\catcode `\^12\catcode `\_12\catcode `\%12\relax}%
\providecommand \@@startlink[1]{}%
\providecommand \@@endlink[0]{}%
\providecommand \url  [0]{\begingroup\@sanitize@url \@url }%
\providecommand \@url [1]{\endgroup\@href {#1}{\urlprefix }}%
\providecommand \urlprefix  [0]{URL }%
\providecommand \Eprint [0]{\href }%
\providecommand \doibase [0]{https://doi.org/}%
\providecommand \selectlanguage [0]{\@gobble}%
\providecommand \bibinfo  [0]{\@secondoftwo}%
\providecommand \bibfield  [0]{\@secondoftwo}%
\providecommand \translation [1]{[#1]}%
\providecommand \BibitemOpen [0]{}%
\providecommand \bibitemStop [0]{}%
\providecommand \bibitemNoStop [0]{.\EOS\space}%
\providecommand \EOS [0]{\spacefactor3000\relax}%
\providecommand \BibitemShut  [1]{\csname bibitem#1\endcsname}%
\let\auto@bib@innerbib\@empty
\bibitem [{\citenamefont {Le}\ \emph {et~al.}(2026)\citenamefont {Le}, \citenamefont {Thareja}, \citenamefont {Konushbaev}, \citenamefont {Pantano}, \citenamefont {Saunderson}, \citenamefont {Phan}, \citenamefont {Mokrousov},\ and\ \citenamefont {Gayles}}]{le2026ultra}%
  \BibitemOpen
  \bibfield  {author} {\bibinfo {author} {\bibfnamefont {D.~K.}\ \bibnamefont {Le}}, \bibinfo {author} {\bibfnamefont {E.}~\bibnamefont {Thareja}}, \bibinfo {author} {\bibfnamefont {B.}~\bibnamefont {Konushbaev}}, \bibinfo {author} {\bibfnamefont {G.}~\bibnamefont {Pantano}}, \bibinfo {author} {\bibfnamefont {T.}~\bibnamefont {Saunderson}}, \bibinfo {author} {\bibfnamefont {M.-H.}\ \bibnamefont {Phan}}, \bibinfo {author} {\bibfnamefont {Y.}~\bibnamefont {Mokrousov}},\ and\ \bibinfo {author} {\bibfnamefont {J.}~\bibnamefont {Gayles}},\ }\bibfield  {title} {\bibinfo {title} {Ultra-stable weyl topology driven by magnetic textures in the shandite compound co3sn2s (2-x) sex},\ }\href@noop {} {\bibfield  {journal} {\bibinfo  {journal} {arXiv preprint arXiv:2601.09922}\ } (\bibinfo {year} {2026})}\BibitemShut {NoStop}%
\bibitem [{\citenamefont {Dominici}\ \emph {et~al.}(2023)\citenamefont {Dominici}, \citenamefont {Rahmani}, \citenamefont {Colas}, \citenamefont {Ballarini}, \citenamefont {De~Giorgi}, \citenamefont {Gigli}, \citenamefont {Sanvitto}, \citenamefont {Laussy},\ and\ \citenamefont {Voronova}}]{dominici2023coupled}%
  \BibitemOpen
  \bibfield  {author} {\bibinfo {author} {\bibfnamefont {L.}~\bibnamefont {Dominici}}, \bibinfo {author} {\bibfnamefont {A.}~\bibnamefont {Rahmani}}, \bibinfo {author} {\bibfnamefont {D.}~\bibnamefont {Colas}}, \bibinfo {author} {\bibfnamefont {D.}~\bibnamefont {Ballarini}}, \bibinfo {author} {\bibfnamefont {M.}~\bibnamefont {De~Giorgi}}, \bibinfo {author} {\bibfnamefont {G.}~\bibnamefont {Gigli}}, \bibinfo {author} {\bibfnamefont {D.}~\bibnamefont {Sanvitto}}, \bibinfo {author} {\bibfnamefont {F.~P.}\ \bibnamefont {Laussy}},\ and\ \bibinfo {author} {\bibfnamefont {N.}~\bibnamefont {Voronova}},\ }\bibfield  {title} {\bibinfo {title} {Coupled quantum vortex kinematics and berry curvature in real space},\ }\href@noop {} {\bibfield  {journal} {\bibinfo  {journal} {Communications Physics}\ }\textbf {\bibinfo {volume} {6}},\ \bibinfo {pages} {197} (\bibinfo {year} {2023})}\BibitemShut {NoStop}%
\bibitem [{\citenamefont {He}\ \emph {et~al.}(2024)\citenamefont {He}, \citenamefont {Yao}, \citenamefont {Pan}, \citenamefont {Arpino}, \citenamefont {Chen}, \citenamefont {Serrano-Sanchez}, \citenamefont {Ju}, \citenamefont {Shi}, \citenamefont {Sun},\ and\ \citenamefont {Felser}}]{he2024enhanced}%
  \BibitemOpen
  \bibfield  {author} {\bibinfo {author} {\bibfnamefont {B.}~\bibnamefont {He}}, \bibinfo {author} {\bibfnamefont {M.}~\bibnamefont {Yao}}, \bibinfo {author} {\bibfnamefont {Y.}~\bibnamefont {Pan}}, \bibinfo {author} {\bibfnamefont {K.~E.}\ \bibnamefont {Arpino}}, \bibinfo {author} {\bibfnamefont {D.}~\bibnamefont {Chen}}, \bibinfo {author} {\bibfnamefont {F.~M.}\ \bibnamefont {Serrano-Sanchez}}, \bibinfo {author} {\bibfnamefont {S.}~\bibnamefont {Ju}}, \bibinfo {author} {\bibfnamefont {M.}~\bibnamefont {Shi}}, \bibinfo {author} {\bibfnamefont {Y.}~\bibnamefont {Sun}},\ and\ \bibinfo {author} {\bibfnamefont {C.}~\bibnamefont {Felser}},\ }\bibfield  {title} {\bibinfo {title} {Enhanced weyl semimetal signature in co3sn2s2 kagome ferromagnet by chlorine doping},\ }\href@noop {} {\bibfield  {journal} {\bibinfo  {journal} {Communications Materials}\ }\textbf {\bibinfo {volume} {5}},\ \bibinfo {pages} {275} (\bibinfo {year} {2024})}\BibitemShut {NoStop}%
\bibitem [{\citenamefont {Raju}\ \emph {et~al.}(2024)\citenamefont {Raju}, \citenamefont {Romero~III}, \citenamefont {Nishio-Hamane}, \citenamefont {Uesugi}, \citenamefont {Asakura}, \citenamefont {Tagay}, \citenamefont {Higo}, \citenamefont {Armitage}, \citenamefont {Broholm},\ and\ \citenamefont {Nakatsuji}}]{raju2024anisotropic}%
  \BibitemOpen
  \bibfield  {author} {\bibinfo {author} {\bibfnamefont {M.}~\bibnamefont {Raju}}, \bibinfo {author} {\bibfnamefont {R.}~\bibnamefont {Romero~III}}, \bibinfo {author} {\bibfnamefont {D.}~\bibnamefont {Nishio-Hamane}}, \bibinfo {author} {\bibfnamefont {R.}~\bibnamefont {Uesugi}}, \bibinfo {author} {\bibfnamefont {M.}~\bibnamefont {Asakura}}, \bibinfo {author} {\bibfnamefont {Z.}~\bibnamefont {Tagay}}, \bibinfo {author} {\bibfnamefont {T.}~\bibnamefont {Higo}}, \bibinfo {author} {\bibfnamefont {N.}~\bibnamefont {Armitage}}, \bibinfo {author} {\bibfnamefont {C.}~\bibnamefont {Broholm}},\ and\ \bibinfo {author} {\bibfnamefont {S.}~\bibnamefont {Nakatsuji}},\ }\bibfield  {title} {\bibinfo {title} {Anisotropic anomalous transport in the kagome-based topological antiferromagnetic mn 3 ga epitaxial thin films},\ }\href@noop {} {\bibfield  {journal} {\bibinfo  {journal} {Physical Review Materials}\ }\textbf {\bibinfo {volume} {8}},\ \bibinfo {pages} {014204} (\bibinfo {year} {2024})}\BibitemShut {NoStop}%
\bibitem [{\citenamefont {Wan}\ \emph {et~al.}(2011)\citenamefont {Wan}, \citenamefont {Turner}, \citenamefont {Vishwanath},\ and\ \citenamefont {Savrasov}}]{wan2011topological}%
  \BibitemOpen
  \bibfield  {author} {\bibinfo {author} {\bibfnamefont {X.}~\bibnamefont {Wan}}, \bibinfo {author} {\bibfnamefont {A.~M.}\ \bibnamefont {Turner}}, \bibinfo {author} {\bibfnamefont {A.}~\bibnamefont {Vishwanath}},\ and\ \bibinfo {author} {\bibfnamefont {S.~Y.}\ \bibnamefont {Savrasov}},\ }\bibfield  {title} {\bibinfo {title} {Topological semimetal and fermi-arc surface states in the electronic structure of pyrochlore iridates},\ }\href@noop {} {\bibfield  {journal} {\bibinfo  {journal} {Physical Review B}\ }\textbf {\bibinfo {volume} {83}},\ \bibinfo {pages} {205101} (\bibinfo {year} {2011})}\BibitemShut {NoStop}%
\bibitem [{\citenamefont {Armitage}\ \emph {et~al.}(2018)\citenamefont {Armitage}, \citenamefont {Mele},\ and\ \citenamefont {Vishwanath}}]{armitage2018weyl}%
  \BibitemOpen
  \bibfield  {author} {\bibinfo {author} {\bibfnamefont {N.}~\bibnamefont {Armitage}}, \bibinfo {author} {\bibfnamefont {E.}~\bibnamefont {Mele}},\ and\ \bibinfo {author} {\bibfnamefont {A.}~\bibnamefont {Vishwanath}},\ }\bibfield  {title} {\bibinfo {title} {Weyl and dirac semimetals in three-dimensional solids},\ }\href@noop {} {\bibfield  {journal} {\bibinfo  {journal} {Reviews of Modern Physics}\ }\textbf {\bibinfo {volume} {90}},\ \bibinfo {pages} {015001} (\bibinfo {year} {2018})}\BibitemShut {NoStop}%
\bibitem [{\citenamefont {Son}\ and\ \citenamefont {Yamamoto}(2012)}]{son2012berry}%
  \BibitemOpen
  \bibfield  {author} {\bibinfo {author} {\bibfnamefont {D.~T.}\ \bibnamefont {Son}}\ and\ \bibinfo {author} {\bibfnamefont {N.}~\bibnamefont {Yamamoto}},\ }\bibfield  {title} {\bibinfo {title} {Berry curvature, triangle anomalies, and the chiral magnetic effect in fermi liquids},\ }\href@noop {} {\bibfield  {journal} {\bibinfo  {journal} {Physical review letters}\ }\textbf {\bibinfo {volume} {109}},\ \bibinfo {pages} {181602} (\bibinfo {year} {2012})}\BibitemShut {NoStop}%
\bibitem [{\citenamefont {Burkov}(2014)}]{burkov2014anomalous}%
  \BibitemOpen
  \bibfield  {author} {\bibinfo {author} {\bibfnamefont {A.}~\bibnamefont {Burkov}},\ }\bibfield  {title} {\bibinfo {title} {Anomalous hall effect in weyl metals},\ }\href@noop {} {\bibfield  {journal} {\bibinfo  {journal} {Physical Review Letters}\ }\textbf {\bibinfo {volume} {113}},\ \bibinfo {pages} {187202} (\bibinfo {year} {2014})}\BibitemShut {NoStop}%
\bibitem [{\citenamefont {Bevan}\ \emph {et~al.}(1997)\citenamefont {Bevan}, \citenamefont {Manninen}, \citenamefont {Cook}, \citenamefont {Hook}, \citenamefont {Hall}, \citenamefont {Vachaspati},\ and\ \citenamefont {Volovik}}]{bevan1997momentum}%
  \BibitemOpen
  \bibfield  {author} {\bibinfo {author} {\bibfnamefont {T.}~\bibnamefont {Bevan}}, \bibinfo {author} {\bibfnamefont {A.}~\bibnamefont {Manninen}}, \bibinfo {author} {\bibfnamefont {J.}~\bibnamefont {Cook}}, \bibinfo {author} {\bibfnamefont {J.}~\bibnamefont {Hook}}, \bibinfo {author} {\bibfnamefont {H.}~\bibnamefont {Hall}}, \bibinfo {author} {\bibfnamefont {T.}~\bibnamefont {Vachaspati}},\ and\ \bibinfo {author} {\bibfnamefont {G.}~\bibnamefont {Volovik}},\ }\bibfield  {title} {\bibinfo {title} {Momentum creation by vortices in superfluid 3he as a model of primordial baryogenesis},\ }\href@noop {} {\bibfield  {journal} {\bibinfo  {journal} {Nature}\ }\textbf {\bibinfo {volume} {386}},\ \bibinfo {pages} {689} (\bibinfo {year} {1997})}\BibitemShut {NoStop}%
\bibitem [{\citenamefont {Volovik}(1999)}]{volovik1999induced}%
  \BibitemOpen
  \bibfield  {author} {\bibinfo {author} {\bibfnamefont {G.}~\bibnamefont {Volovik}},\ }\bibfield  {title} {\bibinfo {title} {On induced cpt-odd chern-simons terms in the 3+ 1 effective action},\ }\href@noop {} {\bibfield  {journal} {\bibinfo  {journal} {Journal of Experimental and Theoretical Physics Letters}\ }\textbf {\bibinfo {volume} {70}},\ \bibinfo {pages} {1} (\bibinfo {year} {1999})}\BibitemShut {NoStop}%
\bibitem [{\citenamefont {Volovik}(2003)}]{volovik2003universe}%
  \BibitemOpen
  \bibfield  {author} {\bibinfo {author} {\bibfnamefont {G.~E.}\ \bibnamefont {Volovik}},\ }\href@noop {} {\emph {\bibinfo {title} {The universe in a helium droplet}}},\ Vol.\ \bibinfo {volume} {117}\ (\bibinfo  {publisher} {Oxford University Press on Demand},\ \bibinfo {year} {2003})\BibitemShut {NoStop}%
\bibitem [{\citenamefont {Novoselov}\ \emph {et~al.}(2006)\citenamefont {Novoselov}, \citenamefont {McCann}, \citenamefont {Morozov}, \citenamefont {Falko}, \citenamefont {Katsnelson}, \citenamefont {Zeitler}, \citenamefont {Jiang}, \citenamefont {Schedin},\ and\ \citenamefont {Geim}}]{novoselov2006unconventional}%
  \BibitemOpen
  \bibfield  {author} {\bibinfo {author} {\bibfnamefont {K.~S.}\ \bibnamefont {Novoselov}}, \bibinfo {author} {\bibfnamefont {E.}~\bibnamefont {McCann}}, \bibinfo {author} {\bibfnamefont {S.}~\bibnamefont {Morozov}}, \bibinfo {author} {\bibfnamefont {V.~I.}\ \bibnamefont {Falko}}, \bibinfo {author} {\bibfnamefont {M.}~\bibnamefont {Katsnelson}}, \bibinfo {author} {\bibfnamefont {U.}~\bibnamefont {Zeitler}}, \bibinfo {author} {\bibfnamefont {D.}~\bibnamefont {Jiang}}, \bibinfo {author} {\bibfnamefont {F.}~\bibnamefont {Schedin}},\ and\ \bibinfo {author} {\bibfnamefont {A.}~\bibnamefont {Geim}},\ }\bibfield  {title} {\bibinfo {title} {Unconventional quantum hall effect and berry’s phase of 2$\pi$ in bilayer graphene},\ }\href@noop {} {\bibfield  {journal} {\bibinfo  {journal} {Nature physics}\ }\textbf {\bibinfo {volume} {2}},\ \bibinfo {pages} {177} (\bibinfo {year} {2006})}\BibitemShut {NoStop}%
\bibitem [{\citenamefont {Berry}(1984)}]{berry1984quantal}%
  \BibitemOpen
  \bibfield  {author} {\bibinfo {author} {\bibfnamefont {M.~V.}\ \bibnamefont {Berry}},\ }\bibfield  {title} {\bibinfo {title} {Quantal phase factors accompanying adiabatic changes},\ }\href@noop {} {\bibfield  {journal} {\bibinfo  {journal} {Proceedings of the Royal Society of London. A. Mathematical and Physical Sciences}\ }\textbf {\bibinfo {volume} {392}},\ \bibinfo {pages} {45} (\bibinfo {year} {1984})}\BibitemShut {NoStop}%
\bibitem [{\citenamefont {Sundaram}\ and\ \citenamefont {Niu}(1999)}]{sundaram1999wave}%
  \BibitemOpen
  \bibfield  {author} {\bibinfo {author} {\bibfnamefont {G.}~\bibnamefont {Sundaram}}\ and\ \bibinfo {author} {\bibfnamefont {Q.}~\bibnamefont {Niu}},\ }\bibfield  {title} {\bibinfo {title} {Wave-packet dynamics in slowly perturbed crystals: Gradient corrections and berry-phase effects},\ }\href@noop {} {\bibfield  {journal} {\bibinfo  {journal} {Physical Review B}\ }\textbf {\bibinfo {volume} {59}},\ \bibinfo {pages} {14915} (\bibinfo {year} {1999})}\BibitemShut {NoStop}%
\bibitem [{\citenamefont {Xiao}\ \emph {et~al.}(2010)\citenamefont {Xiao}, \citenamefont {Chang},\ and\ \citenamefont {Niu}}]{xiao2010berry}%
  \BibitemOpen
  \bibfield  {author} {\bibinfo {author} {\bibfnamefont {D.}~\bibnamefont {Xiao}}, \bibinfo {author} {\bibfnamefont {M.-C.}\ \bibnamefont {Chang}},\ and\ \bibinfo {author} {\bibfnamefont {Q.}~\bibnamefont {Niu}},\ }\bibfield  {title} {\bibinfo {title} {Berry phase effects on electronic properties},\ }\href@noop {} {\bibfield  {journal} {\bibinfo  {journal} {Reviews of modern physics}\ }\textbf {\bibinfo {volume} {82}},\ \bibinfo {pages} {1959} (\bibinfo {year} {2010})}\BibitemShut {NoStop}%
\bibitem [{\citenamefont {Nagaosa}\ and\ \citenamefont {Tokura}(2013)}]{nagaosa2013topological}%
  \BibitemOpen
  \bibfield  {author} {\bibinfo {author} {\bibfnamefont {N.}~\bibnamefont {Nagaosa}}\ and\ \bibinfo {author} {\bibfnamefont {Y.}~\bibnamefont {Tokura}},\ }\bibfield  {title} {\bibinfo {title} {Topological properties and dynamics of magnetic skyrmions},\ }\href@noop {} {\bibfield  {journal} {\bibinfo  {journal} {Nature nanotechnology}\ }\textbf {\bibinfo {volume} {8}},\ \bibinfo {pages} {899} (\bibinfo {year} {2013})}\BibitemShut {NoStop}%
\bibitem [{\citenamefont {Verma}\ \emph {et~al.}(2022)\citenamefont {Verma}, \citenamefont {Addison},\ and\ \citenamefont {Randeria}}]{verma2022unified}%
  \BibitemOpen
  \bibfield  {author} {\bibinfo {author} {\bibfnamefont {N.}~\bibnamefont {Verma}}, \bibinfo {author} {\bibfnamefont {Z.}~\bibnamefont {Addison}},\ and\ \bibinfo {author} {\bibfnamefont {M.}~\bibnamefont {Randeria}},\ }\bibfield  {title} {\bibinfo {title} {Unified theory of the anomalous and topological hall effects with phase-space berry curvatures},\ }\href@noop {} {\bibfield  {journal} {\bibinfo  {journal} {Science Advances}\ }\textbf {\bibinfo {volume} {8}},\ \bibinfo {pages} {eabq2765} (\bibinfo {year} {2022})}\BibitemShut {NoStop}%
\bibitem [{\citenamefont {Goswami}\ and\ \citenamefont {Tewari}(2013)}]{goswami2013axionic}%
  \BibitemOpen
  \bibfield  {author} {\bibinfo {author} {\bibfnamefont {P.}~\bibnamefont {Goswami}}\ and\ \bibinfo {author} {\bibfnamefont {S.}~\bibnamefont {Tewari}},\ }\bibfield  {title} {\bibinfo {title} {Axionic field theory of (3+ 1)-dimensional weyl semimetals},\ }\href@noop {} {\bibfield  {journal} {\bibinfo  {journal} {Physical Review B}\ }\textbf {\bibinfo {volume} {88}},\ \bibinfo {pages} {245107} (\bibinfo {year} {2013})}\BibitemShut {NoStop}%
\bibitem [{\citenamefont {Hirschberger}\ \emph {et~al.}(2019)\citenamefont {Hirschberger}, \citenamefont {Nakajima}, \citenamefont {Gao}, \citenamefont {Peng}, \citenamefont {Kikkawa}, \citenamefont {Kurumaji}, \citenamefont {Kriener}, \citenamefont {Yamasaki}, \citenamefont {Sagayama}, \citenamefont {Nakao} \emph {et~al.}}]{hirschberger2019skyrmion}%
  \BibitemOpen
  \bibfield  {author} {\bibinfo {author} {\bibfnamefont {M.}~\bibnamefont {Hirschberger}}, \bibinfo {author} {\bibfnamefont {T.}~\bibnamefont {Nakajima}}, \bibinfo {author} {\bibfnamefont {S.}~\bibnamefont {Gao}}, \bibinfo {author} {\bibfnamefont {L.}~\bibnamefont {Peng}}, \bibinfo {author} {\bibfnamefont {A.}~\bibnamefont {Kikkawa}}, \bibinfo {author} {\bibfnamefont {T.}~\bibnamefont {Kurumaji}}, \bibinfo {author} {\bibfnamefont {M.}~\bibnamefont {Kriener}}, \bibinfo {author} {\bibfnamefont {Y.}~\bibnamefont {Yamasaki}}, \bibinfo {author} {\bibfnamefont {H.}~\bibnamefont {Sagayama}}, \bibinfo {author} {\bibfnamefont {H.}~\bibnamefont {Nakao}}, \emph {et~al.},\ }\bibfield  {title} {\bibinfo {title} {Skyrmion phase and competing magnetic orders on a breathing kagom{\'e} lattice},\ }\href@noop {} {\bibfield  {journal} {\bibinfo  {journal} {Nature communications}\ }\textbf {\bibinfo {volume} {10}},\ \bibinfo {pages} {5831} (\bibinfo {year} {2019})}\BibitemShut {NoStop}%
\bibitem [{\citenamefont {Hirschberger}\ \emph {et~al.}(2024)\citenamefont {Hirschberger}, \citenamefont {Szigeti}, \citenamefont {Hemmida}, \citenamefont {Hirschmann}, \citenamefont {Esser}, \citenamefont {Ohsumi}, \citenamefont {Tanaka}, \citenamefont {Spitz}, \citenamefont {Gao}, \citenamefont {Kolincio} \emph {et~al.}}]{hirschberger2024lattice}%
  \BibitemOpen
  \bibfield  {author} {\bibinfo {author} {\bibfnamefont {M.}~\bibnamefont {Hirschberger}}, \bibinfo {author} {\bibfnamefont {B.~G.}\ \bibnamefont {Szigeti}}, \bibinfo {author} {\bibfnamefont {M.}~\bibnamefont {Hemmida}}, \bibinfo {author} {\bibfnamefont {M.~M.}\ \bibnamefont {Hirschmann}}, \bibinfo {author} {\bibfnamefont {S.}~\bibnamefont {Esser}}, \bibinfo {author} {\bibfnamefont {H.}~\bibnamefont {Ohsumi}}, \bibinfo {author} {\bibfnamefont {Y.}~\bibnamefont {Tanaka}}, \bibinfo {author} {\bibfnamefont {L.}~\bibnamefont {Spitz}}, \bibinfo {author} {\bibfnamefont {S.}~\bibnamefont {Gao}}, \bibinfo {author} {\bibfnamefont {K.~K.}\ \bibnamefont {Kolincio}}, \emph {et~al.},\ }\bibfield  {title} {\bibinfo {title} {Lattice-commensurate skyrmion texture in a centrosymmetric breathing kagome magnet},\ }\href@noop {} {\bibfield  {journal} {\bibinfo  {journal} {npj Quantum Materials}\ }\textbf {\bibinfo {volume} {9}},\ \bibinfo {pages} {45} (\bibinfo {year} {2024})}\BibitemShut {NoStop}%
\bibitem [{\citenamefont {Du}\ \emph {et~al.}(2020)\citenamefont {Du}, \citenamefont {Han}, \citenamefont {Liu}, \citenamefont {Ren}, \citenamefont {Zhu},\ and\ \citenamefont {Petrovic}}]{du2020room}%
  \BibitemOpen
  \bibfield  {author} {\bibinfo {author} {\bibfnamefont {Q.}~\bibnamefont {Du}}, \bibinfo {author} {\bibfnamefont {M.-G.}\ \bibnamefont {Han}}, \bibinfo {author} {\bibfnamefont {Y.}~\bibnamefont {Liu}}, \bibinfo {author} {\bibfnamefont {W.}~\bibnamefont {Ren}}, \bibinfo {author} {\bibfnamefont {Y.}~\bibnamefont {Zhu}},\ and\ \bibinfo {author} {\bibfnamefont {C.}~\bibnamefont {Petrovic}},\ }\bibfield  {title} {\bibinfo {title} {Room-temperature skyrmion thermopower in fe3sn2},\ }\href@noop {} {\bibfield  {journal} {\bibinfo  {journal} {Advanced Quantum Technologies}\ }\textbf {\bibinfo {volume} {3}},\ \bibinfo {pages} {2000058} (\bibinfo {year} {2020})}\BibitemShut {NoStop}%
\bibitem [{\citenamefont {Vir}\ \emph {et~al.}(2019)\citenamefont {Vir}, \citenamefont {Gayles}, \citenamefont {Sukhanov}, \citenamefont {Kumar}, \citenamefont {Damay}, \citenamefont {Sun}, \citenamefont {K{\"u}bler}, \citenamefont {Shekhar},\ and\ \citenamefont {Felser}}]{vir2019anisotropic}%
  \BibitemOpen
  \bibfield  {author} {\bibinfo {author} {\bibfnamefont {P.}~\bibnamefont {Vir}}, \bibinfo {author} {\bibfnamefont {J.}~\bibnamefont {Gayles}}, \bibinfo {author} {\bibfnamefont {A.}~\bibnamefont {Sukhanov}}, \bibinfo {author} {\bibfnamefont {N.}~\bibnamefont {Kumar}}, \bibinfo {author} {\bibfnamefont {F.}~\bibnamefont {Damay}}, \bibinfo {author} {\bibfnamefont {Y.}~\bibnamefont {Sun}}, \bibinfo {author} {\bibfnamefont {J.}~\bibnamefont {K{\"u}bler}}, \bibinfo {author} {\bibfnamefont {C.}~\bibnamefont {Shekhar}},\ and\ \bibinfo {author} {\bibfnamefont {C.}~\bibnamefont {Felser}},\ }\bibfield  {title} {\bibinfo {title} {Anisotropic topological hall effect with real and momentum space berry curvature in the antiskrymion-hosting heusler compound mn 1.4 ptsn},\ }\href@noop {} {\bibfield  {journal} {\bibinfo  {journal} {Physical Review B}\ }\textbf {\bibinfo {volume} {99}},\ \bibinfo {pages} {140406} (\bibinfo {year} {2019})}\BibitemShut {NoStop}%
\bibitem [{\citenamefont {Addison}\ \emph {et~al.}(2025)\citenamefont {Addison}, \citenamefont {Keyes},\ and\ \citenamefont {Randeria}}]{addison2025anomalous}%
  \BibitemOpen
  \bibfield  {author} {\bibinfo {author} {\bibfnamefont {Z.}~\bibnamefont {Addison}}, \bibinfo {author} {\bibfnamefont {L.}~\bibnamefont {Keyes}},\ and\ \bibinfo {author} {\bibfnamefont {M.}~\bibnamefont {Randeria}},\ }\bibfield  {title} {\bibinfo {title} {Anomalous and topological hall effects with phase-space berry curvatures: Electric, thermal, and thermoelectric transport in magnets},\ }\href@noop {} {\bibfield  {journal} {\bibinfo  {journal} {Physical Review B}\ }\textbf {\bibinfo {volume} {112}},\ \bibinfo {pages} {014446} (\bibinfo {year} {2025})}\BibitemShut {NoStop}%
\bibitem [{\citenamefont {G{\"o}bel}\ \emph {et~al.}(2025)\citenamefont {G{\"o}bel}, \citenamefont {Schimpf},\ and\ \citenamefont {Mertig}}]{gobel2025topological}%
  \BibitemOpen
  \bibfield  {author} {\bibinfo {author} {\bibfnamefont {B.}~\bibnamefont {G{\"o}bel}}, \bibinfo {author} {\bibfnamefont {L.}~\bibnamefont {Schimpf}},\ and\ \bibinfo {author} {\bibfnamefont {I.}~\bibnamefont {Mertig}},\ }\bibfield  {title} {\bibinfo {title} {Topological orbital hall effect caused by skyrmions and antiferromagnetic skyrmions},\ }\href@noop {} {\bibfield  {journal} {\bibinfo  {journal} {Communications Physics}\ }\textbf {\bibinfo {volume} {8}},\ \bibinfo {pages} {17} (\bibinfo {year} {2025})}\BibitemShut {NoStop}%
\bibitem [{\citenamefont {Ahmad}\ and\ \citenamefont {Sharma}(2021)}]{ahmad2021longitudinal}%
  \BibitemOpen
  \bibfield  {author} {\bibinfo {author} {\bibfnamefont {A.}~\bibnamefont {Ahmad}}\ and\ \bibinfo {author} {\bibfnamefont {G.}~\bibnamefont {Sharma}},\ }\bibfield  {title} {\bibinfo {title} {Longitudinal magnetoconductance and the planar hall effect in a lattice model of tilted weyl fermions},\ }\href@noop {} {\bibfield  {journal} {\bibinfo  {journal} {Physical Review B}\ }\textbf {\bibinfo {volume} {103}},\ \bibinfo {pages} {115146} (\bibinfo {year} {2021})}\BibitemShut {NoStop}%
\bibitem [{\citenamefont {Ahmad}\ \emph {et~al.}(2023)\citenamefont {Ahmad}, \citenamefont {Raman}, \citenamefont {Tewari},\ and\ \citenamefont {Sharma}}]{ahmad2023longitudinal}%
  \BibitemOpen
  \bibfield  {author} {\bibinfo {author} {\bibfnamefont {A.}~\bibnamefont {Ahmad}}, \bibinfo {author} {\bibfnamefont {K.~V.}\ \bibnamefont {Raman}}, \bibinfo {author} {\bibfnamefont {S.}~\bibnamefont {Tewari}},\ and\ \bibinfo {author} {\bibfnamefont {G.}~\bibnamefont {Sharma}},\ }\bibfield  {title} {\bibinfo {title} {Longitudinal magnetoconductance and the planar hall conductance in inhomogeneous weyl semimetals},\ }\href@noop {} {\bibfield  {journal} {\bibinfo  {journal} {Physical Review B}\ }\textbf {\bibinfo {volume} {107}},\ \bibinfo {pages} {144206} (\bibinfo {year} {2023})}\BibitemShut {NoStop}%
\bibitem [{\citenamefont {Liu}\ \emph {et~al.}(2013)\citenamefont {Liu}, \citenamefont {Ye},\ and\ \citenamefont {Qi}}]{liu2013chiral}%
  \BibitemOpen
  \bibfield  {author} {\bibinfo {author} {\bibfnamefont {C.-X.}\ \bibnamefont {Liu}}, \bibinfo {author} {\bibfnamefont {P.}~\bibnamefont {Ye}},\ and\ \bibinfo {author} {\bibfnamefont {X.-L.}\ \bibnamefont {Qi}},\ }\bibfield  {title} {\bibinfo {title} {Chiral gauge field and axial anomaly in a weyl semimetal},\ }\href@noop {} {\bibfield  {journal} {\bibinfo  {journal} {Physical Review B}\ }\textbf {\bibinfo {volume} {87}},\ \bibinfo {pages} {235306} (\bibinfo {year} {2013})}\BibitemShut {NoStop}%
\bibitem [{\citenamefont {Grushin}(2012)}]{grushin2012consequences}%
  \BibitemOpen
  \bibfield  {author} {\bibinfo {author} {\bibfnamefont {A.~G.}\ \bibnamefont {Grushin}},\ }\bibfield  {title} {\bibinfo {title} {Consequences of a condensed matter realization of lorentz-violating qed in weyl semi-metals},\ }\href@noop {} {\bibfield  {journal} {\bibinfo  {journal} {Physical Review D}\ }\textbf {\bibinfo {volume} {86}},\ \bibinfo {pages} {045001} (\bibinfo {year} {2012})}\BibitemShut {NoStop}%
\bibitem [{\citenamefont {Zyuzin}\ and\ \citenamefont {Burkov}(2012)}]{zyuzin2012topological}%
  \BibitemOpen
  \bibfield  {author} {\bibinfo {author} {\bibfnamefont {A.}~\bibnamefont {Zyuzin}}\ and\ \bibinfo {author} {\bibfnamefont {A.}~\bibnamefont {Burkov}},\ }\bibfield  {title} {\bibinfo {title} {Topological response in weyl semimetals and the chiral anomaly},\ }\href@noop {} {\bibfield  {journal} {\bibinfo  {journal} {Physical Review B}\ }\textbf {\bibinfo {volume} {86}},\ \bibinfo {pages} {115133} (\bibinfo {year} {2012})}\BibitemShut {NoStop}%
\bibitem [{\citenamefont {Addison}\ \emph {et~al.}(2023)\citenamefont {Addison}, \citenamefont {Keyes},\ and\ \citenamefont {Randeria}}]{addison2023theory}%
  \BibitemOpen
  \bibfield  {author} {\bibinfo {author} {\bibfnamefont {Z.}~\bibnamefont {Addison}}, \bibinfo {author} {\bibfnamefont {L.}~\bibnamefont {Keyes}},\ and\ \bibinfo {author} {\bibfnamefont {M.}~\bibnamefont {Randeria}},\ }\bibfield  {title} {\bibinfo {title} {Theory of topological nernst and thermoelectric transport in chiral magnets},\ }\href@noop {} {\bibfield  {journal} {\bibinfo  {journal} {Physical Review B}\ }\textbf {\bibinfo {volume} {108}},\ \bibinfo {pages} {014419} (\bibinfo {year} {2023})}\BibitemShut {NoStop}%
\bibitem [{\citenamefont {Knoll}\ \emph {et~al.}(2020)\citenamefont {Knoll}, \citenamefont {Timm},\ and\ \citenamefont {Meng}}]{knoll2020negative}%
  \BibitemOpen
  \bibfield  {author} {\bibinfo {author} {\bibfnamefont {A.}~\bibnamefont {Knoll}}, \bibinfo {author} {\bibfnamefont {C.}~\bibnamefont {Timm}},\ and\ \bibinfo {author} {\bibfnamefont {T.}~\bibnamefont {Meng}},\ }\bibfield  {title} {\bibinfo {title} {Negative longitudinal magnetoconductance at weak fields in weyl semimetals},\ }\href@noop {} {\bibfield  {journal} {\bibinfo  {journal} {Physical Review B}\ }\textbf {\bibinfo {volume} {101}},\ \bibinfo {pages} {201402} (\bibinfo {year} {2020})}\BibitemShut {NoStop}%
\bibitem [{\citenamefont {Sharma}\ \emph {et~al.}(2017{\natexlab{a}})\citenamefont {Sharma}, \citenamefont {Goswami},\ and\ \citenamefont {Tewari}}]{sharma2017chiral}%
  \BibitemOpen
  \bibfield  {author} {\bibinfo {author} {\bibfnamefont {G.}~\bibnamefont {Sharma}}, \bibinfo {author} {\bibfnamefont {P.}~\bibnamefont {Goswami}},\ and\ \bibinfo {author} {\bibfnamefont {S.}~\bibnamefont {Tewari}},\ }\bibfield  {title} {\bibinfo {title} {Chiral anomaly and longitudinal magnetotransport in type-ii weyl semimetals},\ }\href@noop {} {\bibfield  {journal} {\bibinfo  {journal} {Physical Review B}\ }\textbf {\bibinfo {volume} {96}},\ \bibinfo {pages} {045112} (\bibinfo {year} {2017}{\natexlab{a}})}\BibitemShut {NoStop}%
\bibitem [{\citenamefont {Sharma}\ \emph {et~al.}(2020)\citenamefont {Sharma}, \citenamefont {Nandy},\ and\ \citenamefont {Tewari}}]{sharma2020sign}%
  \BibitemOpen
  \bibfield  {author} {\bibinfo {author} {\bibfnamefont {G.}~\bibnamefont {Sharma}}, \bibinfo {author} {\bibfnamefont {S.}~\bibnamefont {Nandy}},\ and\ \bibinfo {author} {\bibfnamefont {S.}~\bibnamefont {Tewari}},\ }\bibfield  {title} {\bibinfo {title} {Sign of longitudinal magnetoconductivity and the planar hall effect in weyl semimetals},\ }\href@noop {} {\bibfield  {journal} {\bibinfo  {journal} {Physical Review B}\ }\textbf {\bibinfo {volume} {102}},\ \bibinfo {pages} {205107} (\bibinfo {year} {2020})}\BibitemShut {NoStop}%
\bibitem [{\citenamefont {Ahmad}\ and\ \citenamefont {Sharma}(2024)}]{ahmad2024chiral}%
  \BibitemOpen
  \bibfield  {author} {\bibinfo {author} {\bibfnamefont {A.}~\bibnamefont {Ahmad}}\ and\ \bibinfo {author} {\bibfnamefont {G.}~\bibnamefont {Sharma}},\ }\bibfield  {title} {\bibinfo {title} {Chiral anomaly and longitudinal magnetoconductance in pseudospin-1 fermions},\ }\href@noop {} {\bibfield  {journal} {\bibinfo  {journal} {arXiv preprint arXiv:2412.10500}\ } (\bibinfo {year} {2024})}\BibitemShut {NoStop}%
\bibitem [{\citenamefont {Ahmad}\ \emph {et~al.}(2024)\citenamefont {Ahmad}, \citenamefont {Varma},\ and\ \citenamefont {Sharma}}]{ahmad2024geometry}%
  \BibitemOpen
  \bibfield  {author} {\bibinfo {author} {\bibfnamefont {A.}~\bibnamefont {Ahmad}}, \bibinfo {author} {\bibfnamefont {G.}~\bibnamefont {Varma}},\ and\ \bibinfo {author} {\bibfnamefont {G.}~\bibnamefont {Sharma}},\ }\bibfield  {title} {\bibinfo {title} {Geometry, anomaly, topology, and transport in weyl fermions},\ }\href@noop {} {\bibfield  {journal} {\bibinfo  {journal} {Journal of Physics: Condensed Matter}\ }\textbf {\bibinfo {volume} {37}},\ \bibinfo {pages} {043001} (\bibinfo {year} {2024})}\BibitemShut {NoStop}%
\bibitem [{\citenamefont {Ahmad}\ \emph {et~al.}(2025)\citenamefont {Ahmad}, \citenamefont {K},\ and\ \citenamefont {Sharma}}]{ahmad2025chiral}%
  \BibitemOpen
  \bibfield  {author} {\bibinfo {author} {\bibfnamefont {A.}~\bibnamefont {Ahmad}}, \bibinfo {author} {\bibfnamefont {G.~V.}\ \bibnamefont {K}},\ and\ \bibinfo {author} {\bibfnamefont {G.}~\bibnamefont {Sharma}},\ }\bibfield  {title} {\bibinfo {title} {Chiral anomaly induced nonlinear hall effect in three-dimensional chiral fermions},\ }\href@noop {} {\bibfield  {journal} {\bibinfo  {journal} {Physical Review B}\ }\textbf {\bibinfo {volume} {111}},\ \bibinfo {pages} {035138} (\bibinfo {year} {2025})}\BibitemShut {NoStop}%
\bibitem [{\citenamefont {Mandal}(2025{\natexlab{a}})}]{mandal2025chiral}%
  \BibitemOpen
  \bibfield  {author} {\bibinfo {author} {\bibfnamefont {I.}~\bibnamefont {Mandal}},\ }\bibfield  {title} {\bibinfo {title} {Chiral anomaly and internode scatterings in multifold semimetals},\ }\href@noop {} {\bibfield  {journal} {\bibinfo  {journal} {Physical Review B}\ }\textbf {\bibinfo {volume} {111}},\ \bibinfo {pages} {165116} (\bibinfo {year} {2025}{\natexlab{a}})}\BibitemShut {NoStop}%
\bibitem [{\citenamefont {Mandal}(2025{\natexlab{b}})}]{mandal2025disentangling}%
  \BibitemOpen
  \bibfield  {author} {\bibinfo {author} {\bibfnamefont {I.}~\bibnamefont {Mandal}},\ }\bibfield  {title} {\bibinfo {title} {Disentangling contributions to longitudinal magnetoconductivity for kramers-weyl nodes},\ }\href@noop {} {\bibfield  {journal} {\bibinfo  {journal} {arXiv preprint arXiv:2506.07913}\ } (\bibinfo {year} {2025}{\natexlab{b}})}\BibitemShut {NoStop}%
\bibitem [{\citenamefont {Ghosh}\ \emph {et~al.}(2024)\citenamefont {Ghosh}, \citenamefont {Haidar},\ and\ \citenamefont {Mandal}}]{ghosh2024linear}%
  \BibitemOpen
  \bibfield  {author} {\bibinfo {author} {\bibfnamefont {R.}~\bibnamefont {Ghosh}}, \bibinfo {author} {\bibfnamefont {F.}~\bibnamefont {Haidar}},\ and\ \bibinfo {author} {\bibfnamefont {I.}~\bibnamefont {Mandal}},\ }\bibfield  {title} {\bibinfo {title} {Linear response in planar hall and thermal hall setups for rarita-schwinger-weyl semimetals},\ }\href@noop {} {\bibfield  {journal} {\bibinfo  {journal} {Physical Review B}\ }\textbf {\bibinfo {volume} {110}},\ \bibinfo {pages} {245113} (\bibinfo {year} {2024})}\BibitemShut {NoStop}%
\bibitem [{\citenamefont {Mandal}(2026{\natexlab{a}})}]{mandal2026longitudinal}%
  \BibitemOpen
  \bibfield  {author} {\bibinfo {author} {\bibfnamefont {I.}~\bibnamefont {Mandal}},\ }\bibfield  {title} {\bibinfo {title} {Longitudinal magnetoconductivity in multifold semimetals exemplified by pseudospin-1 nodal points},\ }\href@noop {} {\bibfield  {journal} {\bibinfo  {journal} {The European Physical Journal Plus}\ }\textbf {\bibinfo {volume} {141}},\ \bibinfo {pages} {75} (\bibinfo {year} {2026}{\natexlab{a}})}\BibitemShut {NoStop}%
\bibitem [{\citenamefont {Mandal}({\natexlab{a}})}]{mandal2506disentangling}%
  \BibitemOpen
  \bibfield  {author} {\bibinfo {author} {\bibfnamefont {I.}~\bibnamefont {Mandal}},\ }\bibfield  {title} {\bibinfo {title} {Disentangling contributions to longitudinal magnetoconductivity for kramers-weyl nodes, arxiv e-prints (2025)},\ }\href@noop {} {\bibfield  {journal} {\bibinfo  {journal} {arXiv preprint arXiv:2506.07913}\ } ({\natexlab{a}})}\BibitemShut {NoStop}%
\bibitem [{\citenamefont {Mandal}({\natexlab{b}})}]{mandal2506distinguishing}%
  \BibitemOpen
  \bibfield  {author} {\bibinfo {author} {\bibfnamefont {I.}~\bibnamefont {Mandal}},\ }\bibfield  {title} {\bibinfo {title} {Distinguishing features of longitudinal magnetoconductivity for a rarita-schwinger-weyl node, arxiv e-prints (2025b)},\ }\href@noop {} {\bibfield  {journal} {\bibinfo  {journal} {arXiv preprint arXiv:2506.12380}\ } ({\natexlab{b}})}\BibitemShut {NoStop}%
\bibitem [{\citenamefont {Mandal}(2026{\natexlab{b}})}]{mandal2026unconventional}%
  \BibitemOpen
  \bibfield  {author} {\bibinfo {author} {\bibfnamefont {I.}~\bibnamefont {Mandal}},\ }\bibfield  {title} {\bibinfo {title} {Unconventional magnetoelectric conductivity and electrochemical response from dipole-like sources of berry curvature, arxiv e-prints 10.48550},\ }\href@noop {} {\bibfield  {journal} {\bibinfo  {journal} {arXiv preprint arXiv.2602.08844}\ } (\bibinfo {year} {2026}{\natexlab{b}})}\BibitemShut {NoStop}%
\bibitem [{\citenamefont {Yan}\ and\ \citenamefont {Felser}(2017)}]{yan2017topological}%
  \BibitemOpen
  \bibfield  {author} {\bibinfo {author} {\bibfnamefont {B.}~\bibnamefont {Yan}}\ and\ \bibinfo {author} {\bibfnamefont {C.}~\bibnamefont {Felser}},\ }\bibfield  {title} {\bibinfo {title} {Topological materials: Weyl semimetals},\ }\href@noop {} {\bibfield  {journal} {\bibinfo  {journal} {Annual Review of Condensed Matter Physics}\ }\textbf {\bibinfo {volume} {8}},\ \bibinfo {pages} {337} (\bibinfo {year} {2017})}\BibitemShut {NoStop}%
\bibitem [{\citenamefont {Lv}\ \emph {et~al.}(2015)\citenamefont {Lv}, \citenamefont {Weng}, \citenamefont {Fu}, \citenamefont {Wang}, \citenamefont {Miao}, \citenamefont {Ma}, \citenamefont {Richard}, \citenamefont {Huang}, \citenamefont {Zhao}, \citenamefont {Chen} \emph {et~al.}}]{lv2015experimental}%
  \BibitemOpen
  \bibfield  {author} {\bibinfo {author} {\bibfnamefont {B.}~\bibnamefont {Lv}}, \bibinfo {author} {\bibfnamefont {H.}~\bibnamefont {Weng}}, \bibinfo {author} {\bibfnamefont {B.}~\bibnamefont {Fu}}, \bibinfo {author} {\bibfnamefont {X.~P.}\ \bibnamefont {Wang}}, \bibinfo {author} {\bibfnamefont {H.}~\bibnamefont {Miao}}, \bibinfo {author} {\bibfnamefont {J.}~\bibnamefont {Ma}}, \bibinfo {author} {\bibfnamefont {P.}~\bibnamefont {Richard}}, \bibinfo {author} {\bibfnamefont {X.}~\bibnamefont {Huang}}, \bibinfo {author} {\bibfnamefont {L.}~\bibnamefont {Zhao}}, \bibinfo {author} {\bibfnamefont {G.}~\bibnamefont {Chen}}, \emph {et~al.},\ }\bibfield  {title} {\bibinfo {title} {Experimental discovery of weyl semimetal taas},\ }\href@noop {} {\bibfield  {journal} {\bibinfo  {journal} {Physical Review X}\ }\textbf {\bibinfo {volume} {5}},\ \bibinfo {pages} {031013} (\bibinfo {year} {2015})}\BibitemShut {NoStop}%
\bibitem [{\citenamefont {Hasan}\ \emph {et~al.}(2021)\citenamefont {Hasan}, \citenamefont {Chang}, \citenamefont {Belopolski}, \citenamefont {Bian}, \citenamefont {Xu},\ and\ \citenamefont {Yin}}]{hasan2021weyl}%
  \BibitemOpen
  \bibfield  {author} {\bibinfo {author} {\bibfnamefont {M.~Z.}\ \bibnamefont {Hasan}}, \bibinfo {author} {\bibfnamefont {G.}~\bibnamefont {Chang}}, \bibinfo {author} {\bibfnamefont {I.}~\bibnamefont {Belopolski}}, \bibinfo {author} {\bibfnamefont {G.}~\bibnamefont {Bian}}, \bibinfo {author} {\bibfnamefont {S.-Y.}\ \bibnamefont {Xu}},\ and\ \bibinfo {author} {\bibfnamefont {J.-X.}\ \bibnamefont {Yin}},\ }\bibfield  {title} {\bibinfo {title} {Weyl, dirac and high-fold chiral fermions in topological quantum matter},\ }\href@noop {} {\bibfield  {journal} {\bibinfo  {journal} {Nature Reviews Materials}\ }\textbf {\bibinfo {volume} {6}},\ \bibinfo {pages} {784} (\bibinfo {year} {2021})}\BibitemShut {NoStop}%
\bibitem [{\citenamefont {Varma}\ \emph {et~al.}(2026)\citenamefont {Varma}, \citenamefont {Raza},\ and\ \citenamefont {Ahmad}}]{varma2026chiral}%
  \BibitemOpen
  \bibfield  {author} {\bibinfo {author} {\bibfnamefont {G.}~\bibnamefont {Varma}}, \bibinfo {author} {\bibfnamefont {M.~H.}\ \bibnamefont {Raza}},\ and\ \bibinfo {author} {\bibfnamefont {A.}~\bibnamefont {Ahmad}},\ }\bibfield  {title} {\bibinfo {title} {Chiral anomaly-induced nonlinear hall effect in spin-orbit coupled noncentrosymmetric metals},\ }\href@noop {} {\bibfield  {journal} {\bibinfo  {journal} {Physical Review B}\ }\textbf {\bibinfo {volume} {113}},\ \bibinfo {pages} {035112} (\bibinfo {year} {2026})}\BibitemShut {NoStop}%
\bibitem [{\citenamefont {Ahmad}\ and\ \citenamefont {Sharma}(2025)}]{ahmad2025longitudinal}%
  \BibitemOpen
  \bibfield  {author} {\bibinfo {author} {\bibfnamefont {A.}~\bibnamefont {Ahmad}}\ and\ \bibinfo {author} {\bibfnamefont {G.}~\bibnamefont {Sharma}},\ }\bibfield  {title} {\bibinfo {title} {Longitudinal magnetoconductance of higher-pseudospin fermions},\ }\href@noop {} {\bibfield  {journal} {\bibinfo  {journal} {Physical Review B}\ }\textbf {\bibinfo {volume} {112}},\ \bibinfo {pages} {045135} (\bibinfo {year} {2025})}\BibitemShut {NoStop}%
\bibitem [{\citenamefont {Nandy}\ \emph {et~al.}(2017)\citenamefont {Nandy}, \citenamefont {Sharma}, \citenamefont {Taraphder},\ and\ \citenamefont {Tewari}}]{nandy2017chiral}%
  \BibitemOpen
  \bibfield  {author} {\bibinfo {author} {\bibfnamefont {S.}~\bibnamefont {Nandy}}, \bibinfo {author} {\bibfnamefont {G.}~\bibnamefont {Sharma}}, \bibinfo {author} {\bibfnamefont {A.}~\bibnamefont {Taraphder}},\ and\ \bibinfo {author} {\bibfnamefont {S.}~\bibnamefont {Tewari}},\ }\bibfield  {title} {\bibinfo {title} {Chiral anomaly as the origin of the planar hall effect in weyl semimetals},\ }\href@noop {} {\bibfield  {journal} {\bibinfo  {journal} {Physical review letters}\ }\textbf {\bibinfo {volume} {119}},\ \bibinfo {pages} {176804} (\bibinfo {year} {2017})}\BibitemShut {NoStop}%
\bibitem [{\citenamefont {Nandy}\ \emph {et~al.}(2021{\natexlab{a}})\citenamefont {Nandy}, \citenamefont {Zeng},\ and\ \citenamefont {Tewari}}]{nandy2021chiral}%
  \BibitemOpen
  \bibfield  {author} {\bibinfo {author} {\bibfnamefont {S.}~\bibnamefont {Nandy}}, \bibinfo {author} {\bibfnamefont {C.}~\bibnamefont {Zeng}},\ and\ \bibinfo {author} {\bibfnamefont {S.}~\bibnamefont {Tewari}},\ }\bibfield  {title} {\bibinfo {title} {Chiral anomaly induced nonlinear hall effect in semimetals with multiple weyl points},\ }\href@noop {} {\bibfield  {journal} {\bibinfo  {journal} {Physical Review B}\ }\textbf {\bibinfo {volume} {104}},\ \bibinfo {pages} {205124} (\bibinfo {year} {2021}{\natexlab{a}})}\BibitemShut {NoStop}%
\bibitem [{\citenamefont {Zhang}\ \emph {et~al.}(2016)\citenamefont {Zhang}, \citenamefont {Lu},\ and\ \citenamefont {Shen}}]{zhang2016linear}%
  \BibitemOpen
  \bibfield  {author} {\bibinfo {author} {\bibfnamefont {S.-B.}\ \bibnamefont {Zhang}}, \bibinfo {author} {\bibfnamefont {H.-Z.}\ \bibnamefont {Lu}},\ and\ \bibinfo {author} {\bibfnamefont {S.-Q.}\ \bibnamefont {Shen}},\ }\bibfield  {title} {\bibinfo {title} {Linear magnetoconductivity in an intrinsic topological weyl semimetal},\ }\href@noop {} {\bibfield  {journal} {\bibinfo  {journal} {New Journal of Physics}\ }\textbf {\bibinfo {volume} {18}},\ \bibinfo {pages} {053039} (\bibinfo {year} {2016})}\BibitemShut {NoStop}%
\bibitem [{\citenamefont {Yang}\ \emph {et~al.}(2015)\citenamefont {Yang}, \citenamefont {Pan},\ and\ \citenamefont {Zhang}}]{yang2015chirality}%
  \BibitemOpen
  \bibfield  {author} {\bibinfo {author} {\bibfnamefont {S.~A.}\ \bibnamefont {Yang}}, \bibinfo {author} {\bibfnamefont {H.}~\bibnamefont {Pan}},\ and\ \bibinfo {author} {\bibfnamefont {F.}~\bibnamefont {Zhang}},\ }\bibfield  {title} {\bibinfo {title} {Chirality-dependent hall effect in weyl semimetals},\ }\href@noop {} {\bibfield  {journal} {\bibinfo  {journal} {Physical review letters}\ }\textbf {\bibinfo {volume} {115}},\ \bibinfo {pages} {156603} (\bibinfo {year} {2015})}\BibitemShut {NoStop}%
\bibitem [{\citenamefont {Nandy}\ \emph {et~al.}(2021{\natexlab{b}})\citenamefont {Nandy}, \citenamefont {Zeng},\ and\ \citenamefont {Tewari}}]{PhysRevB.104.205124}%
  \BibitemOpen
  \bibfield  {author} {\bibinfo {author} {\bibfnamefont {S.}~\bibnamefont {Nandy}}, \bibinfo {author} {\bibfnamefont {C.}~\bibnamefont {Zeng}},\ and\ \bibinfo {author} {\bibfnamefont {S.}~\bibnamefont {Tewari}},\ }\bibfield  {title} {\bibinfo {title} {Chiral anomaly induced nonlinear hall effect in semimetals with multiple weyl points},\ }\href {https://doi.org/10.1103/PhysRevB.104.205124} {\bibfield  {journal} {\bibinfo  {journal} {Phys. Rev. B}\ }\textbf {\bibinfo {volume} {104}},\ \bibinfo {pages} {205124} (\bibinfo {year} {2021}{\natexlab{b}})}\BibitemShut {NoStop}%
\bibitem [{\citenamefont {Das}\ and\ \citenamefont {Agarwal}(2019{\natexlab{a}})}]{das2019berry}%
  \BibitemOpen
  \bibfield  {author} {\bibinfo {author} {\bibfnamefont {K.}~\bibnamefont {Das}}\ and\ \bibinfo {author} {\bibfnamefont {A.}~\bibnamefont {Agarwal}},\ }\bibfield  {title} {\bibinfo {title} {Berry curvature induced thermopower in type-i and type-ii weyl semimetals},\ }\href@noop {} {\bibfield  {journal} {\bibinfo  {journal} {Physical Review B}\ }\textbf {\bibinfo {volume} {100}},\ \bibinfo {pages} {085406} (\bibinfo {year} {2019}{\natexlab{a}})}\BibitemShut {NoStop}%
\bibitem [{\citenamefont {Das}\ \emph {et~al.}(2020)\citenamefont {Das}, \citenamefont {Singh},\ and\ \citenamefont {Agarwal}}]{das2020chiral}%
  \BibitemOpen
  \bibfield  {author} {\bibinfo {author} {\bibfnamefont {K.}~\bibnamefont {Das}}, \bibinfo {author} {\bibfnamefont {S.~K.}\ \bibnamefont {Singh}},\ and\ \bibinfo {author} {\bibfnamefont {A.}~\bibnamefont {Agarwal}},\ }\bibfield  {title} {\bibinfo {title} {Chiral anomalies induced transport in weyl metals in quantizing magnetic field},\ }\href@noop {} {\bibfield  {journal} {\bibinfo  {journal} {Physical Review Research}\ }\textbf {\bibinfo {volume} {2}},\ \bibinfo {pages} {033511} (\bibinfo {year} {2020})}\BibitemShut {NoStop}%
\bibitem [{\citenamefont {Das}\ \emph {et~al.}(2022)\citenamefont {Das}, \citenamefont {Das},\ and\ \citenamefont {Agarwal}}]{das2022nonlinear}%
  \BibitemOpen
  \bibfield  {author} {\bibinfo {author} {\bibfnamefont {S.}~\bibnamefont {Das}}, \bibinfo {author} {\bibfnamefont {K.}~\bibnamefont {Das}},\ and\ \bibinfo {author} {\bibfnamefont {A.}~\bibnamefont {Agarwal}},\ }\bibfield  {title} {\bibinfo {title} {Nonlinear magnetoconductivity in weyl and multi-weyl semimetals in quantizing magnetic field},\ }\href@noop {} {\bibfield  {journal} {\bibinfo  {journal} {Physical Review B}\ }\textbf {\bibinfo {volume} {105}},\ \bibinfo {pages} {235408} (\bibinfo {year} {2022})}\BibitemShut {NoStop}%
\bibitem [{\citenamefont {Sharma}\ and\ \citenamefont {Tewari}(2019)}]{sharma2019transverse}%
  \BibitemOpen
  \bibfield  {author} {\bibinfo {author} {\bibfnamefont {G.}~\bibnamefont {Sharma}}\ and\ \bibinfo {author} {\bibfnamefont {S.}~\bibnamefont {Tewari}},\ }\bibfield  {title} {\bibinfo {title} {Transverse thermopower in dirac and weyl semimetals},\ }\href@noop {} {\bibfield  {journal} {\bibinfo  {journal} {Physical Review B}\ }\textbf {\bibinfo {volume} {100}},\ \bibinfo {pages} {195113} (\bibinfo {year} {2019})}\BibitemShut {NoStop}%
\bibitem [{\citenamefont {Sharma}\ \emph {et~al.}(2016)\citenamefont {Sharma}, \citenamefont {Goswami},\ and\ \citenamefont {Tewari}}]{sharma2016nernst}%
  \BibitemOpen
  \bibfield  {author} {\bibinfo {author} {\bibfnamefont {G.}~\bibnamefont {Sharma}}, \bibinfo {author} {\bibfnamefont {P.}~\bibnamefont {Goswami}},\ and\ \bibinfo {author} {\bibfnamefont {S.}~\bibnamefont {Tewari}},\ }\bibfield  {title} {\bibinfo {title} {Nernst and magnetothermal conductivity in a lattice model of weyl fermions},\ }\href@noop {} {\bibfield  {journal} {\bibinfo  {journal} {Physical Review B}\ }\textbf {\bibinfo {volume} {93}},\ \bibinfo {pages} {035116} (\bibinfo {year} {2016})}\BibitemShut {NoStop}%
\bibitem [{\citenamefont {Zeng}\ \emph {et~al.}(2021)\citenamefont {Zeng}, \citenamefont {Nandy},\ and\ \citenamefont {Tewari}}]{zeng2021nonlinear}%
  \BibitemOpen
  \bibfield  {author} {\bibinfo {author} {\bibfnamefont {C.}~\bibnamefont {Zeng}}, \bibinfo {author} {\bibfnamefont {S.}~\bibnamefont {Nandy}},\ and\ \bibinfo {author} {\bibfnamefont {S.}~\bibnamefont {Tewari}},\ }\bibfield  {title} {\bibinfo {title} {Nonlinear transport in weyl semimetals induced by berry curvature dipole},\ }\href@noop {} {\bibfield  {journal} {\bibinfo  {journal} {Physical Review B}\ }\textbf {\bibinfo {volume} {103}},\ \bibinfo {pages} {245119} (\bibinfo {year} {2021})}\BibitemShut {NoStop}%
\bibitem [{\citenamefont {Sharma}\ \emph {et~al.}(2017{\natexlab{b}})\citenamefont {Sharma}, \citenamefont {Moore}, \citenamefont {Saha},\ and\ \citenamefont {Tewari}}]{sharma2017nernst}%
  \BibitemOpen
  \bibfield  {author} {\bibinfo {author} {\bibfnamefont {G.}~\bibnamefont {Sharma}}, \bibinfo {author} {\bibfnamefont {C.}~\bibnamefont {Moore}}, \bibinfo {author} {\bibfnamefont {S.}~\bibnamefont {Saha}},\ and\ \bibinfo {author} {\bibfnamefont {S.}~\bibnamefont {Tewari}},\ }\bibfield  {title} {\bibinfo {title} {Nernst effect in dirac and inversion-asymmetric weyl semimetals},\ }\href@noop {} {\bibfield  {journal} {\bibinfo  {journal} {Physical Review B}\ }\textbf {\bibinfo {volume} {96}},\ \bibinfo {pages} {195119} (\bibinfo {year} {2017}{\natexlab{b}})}\BibitemShut {NoStop}%
\bibitem [{\citenamefont {Nandy}\ \emph {et~al.}(2019)\citenamefont {Nandy}, \citenamefont {Taraphder},\ and\ \citenamefont {Tewari}}]{nandy2019planar}%
  \BibitemOpen
  \bibfield  {author} {\bibinfo {author} {\bibfnamefont {S.}~\bibnamefont {Nandy}}, \bibinfo {author} {\bibfnamefont {A.}~\bibnamefont {Taraphder}},\ and\ \bibinfo {author} {\bibfnamefont {S.}~\bibnamefont {Tewari}},\ }\bibfield  {title} {\bibinfo {title} {Planar thermal hall effect in weyl semimetals},\ }\href@noop {} {\bibfield  {journal} {\bibinfo  {journal} {Physical Review B}\ }\textbf {\bibinfo {volume} {100}},\ \bibinfo {pages} {115139} (\bibinfo {year} {2019})}\BibitemShut {NoStop}%
\bibitem [{\citenamefont {Das}\ and\ \citenamefont {Agarwal}(2020)}]{das2020thermal}%
  \BibitemOpen
  \bibfield  {author} {\bibinfo {author} {\bibfnamefont {K.}~\bibnamefont {Das}}\ and\ \bibinfo {author} {\bibfnamefont {A.}~\bibnamefont {Agarwal}},\ }\bibfield  {title} {\bibinfo {title} {Thermal and gravitational chiral anomaly induced magneto-transport in weyl semimetals},\ }\href@noop {} {\bibfield  {journal} {\bibinfo  {journal} {Physical Review Research}\ }\textbf {\bibinfo {volume} {2}},\ \bibinfo {pages} {013088} (\bibinfo {year} {2020})}\BibitemShut {NoStop}%
\bibitem [{\citenamefont {Das}\ \emph {et~al.}(2023)\citenamefont {Das}, \citenamefont {Das},\ and\ \citenamefont {Agarwal}}]{das2023chiral}%
  \BibitemOpen
  \bibfield  {author} {\bibinfo {author} {\bibfnamefont {S.}~\bibnamefont {Das}}, \bibinfo {author} {\bibfnamefont {K.}~\bibnamefont {Das}},\ and\ \bibinfo {author} {\bibfnamefont {A.}~\bibnamefont {Agarwal}},\ }\bibfield  {title} {\bibinfo {title} {Chiral anomalies in three-dimensional spin-orbit coupled metals: Electrical, thermal, and gravitational anomalies},\ }\href@noop {} {\bibfield  {journal} {\bibinfo  {journal} {Physical Review B}\ }\textbf {\bibinfo {volume} {108}},\ \bibinfo {pages} {045405} (\bibinfo {year} {2023})}\BibitemShut {NoStop}%
\bibitem [{\citenamefont {Mandal}\ \emph {et~al.}(2022)\citenamefont {Mandal}, \citenamefont {Das},\ and\ \citenamefont {Agarwal}}]{mandal2022chiral}%
  \BibitemOpen
  \bibfield  {author} {\bibinfo {author} {\bibfnamefont {D.}~\bibnamefont {Mandal}}, \bibinfo {author} {\bibfnamefont {K.}~\bibnamefont {Das}},\ and\ \bibinfo {author} {\bibfnamefont {A.}~\bibnamefont {Agarwal}},\ }\bibfield  {title} {\bibinfo {title} {Chiral anomaly and nonlinear magnetotransport in time reversal symmetric weyl semimetals},\ }\href@noop {} {\bibfield  {journal} {\bibinfo  {journal} {Physical Review B}\ }\textbf {\bibinfo {volume} {106}},\ \bibinfo {pages} {035423} (\bibinfo {year} {2022})}\BibitemShut {NoStop}%
\bibitem [{\citenamefont {Mandal}\ and\ \citenamefont {Saha}(2020)}]{mandal2020thermopower}%
  \BibitemOpen
  \bibfield  {author} {\bibinfo {author} {\bibfnamefont {I.}~\bibnamefont {Mandal}}\ and\ \bibinfo {author} {\bibfnamefont {K.}~\bibnamefont {Saha}},\ }\bibfield  {title} {\bibinfo {title} {Thermopower in an anisotropic two-dimensional weyl semimetal},\ }\href@noop {} {\bibfield  {journal} {\bibinfo  {journal} {Physical Review B}\ }\textbf {\bibinfo {volume} {101}},\ \bibinfo {pages} {045101} (\bibinfo {year} {2020})}\BibitemShut {NoStop}%
\bibitem [{\citenamefont {Mandal}\ and\ \citenamefont {Sen}(2021)}]{mandal2021tunneling}%
  \BibitemOpen
  \bibfield  {author} {\bibinfo {author} {\bibfnamefont {I.}~\bibnamefont {Mandal}}\ and\ \bibinfo {author} {\bibfnamefont {A.}~\bibnamefont {Sen}},\ }\bibfield  {title} {\bibinfo {title} {Tunneling of multi-weyl semimetals through a potential barrier under the influence of magnetic fields},\ }\href@noop {} {\bibfield  {journal} {\bibinfo  {journal} {Physics Letters A}\ }\textbf {\bibinfo {volume} {399}},\ \bibinfo {pages} {127293} (\bibinfo {year} {2021})}\BibitemShut {NoStop}%
\bibitem [{\citenamefont {Schulz}\ \emph {et~al.}(2012)\citenamefont {Schulz}, \citenamefont {Ritz}, \citenamefont {Bauer}, \citenamefont {Halder}, \citenamefont {Wagner}, \citenamefont {Franz}, \citenamefont {Pfleiderer}, \citenamefont {Everschor}, \citenamefont {Garst},\ and\ \citenamefont {Rosch}}]{schulz2012emergent}%
  \BibitemOpen
  \bibfield  {author} {\bibinfo {author} {\bibfnamefont {T.}~\bibnamefont {Schulz}}, \bibinfo {author} {\bibfnamefont {R.}~\bibnamefont {Ritz}}, \bibinfo {author} {\bibfnamefont {A.}~\bibnamefont {Bauer}}, \bibinfo {author} {\bibfnamefont {M.}~\bibnamefont {Halder}}, \bibinfo {author} {\bibfnamefont {M.}~\bibnamefont {Wagner}}, \bibinfo {author} {\bibfnamefont {C.}~\bibnamefont {Franz}}, \bibinfo {author} {\bibfnamefont {C.}~\bibnamefont {Pfleiderer}}, \bibinfo {author} {\bibfnamefont {K.}~\bibnamefont {Everschor}}, \bibinfo {author} {\bibfnamefont {M.}~\bibnamefont {Garst}},\ and\ \bibinfo {author} {\bibfnamefont {A.}~\bibnamefont {Rosch}},\ }\bibfield  {title} {\bibinfo {title} {Emergent electrodynamics of skyrmions in a chiral magnet},\ }\href@noop {} {\bibfield  {journal} {\bibinfo  {journal} {Nature Physics}\ }\textbf {\bibinfo {volume} {8}},\ \bibinfo {pages} {301} (\bibinfo {year} {2012})}\BibitemShut {NoStop}%
\bibitem [{\citenamefont {Tokura}\ and\ \citenamefont {Kanazawa}(2020)}]{tokura2020magnetic}%
  \BibitemOpen
  \bibfield  {author} {\bibinfo {author} {\bibfnamefont {Y.}~\bibnamefont {Tokura}}\ and\ \bibinfo {author} {\bibfnamefont {N.}~\bibnamefont {Kanazawa}},\ }\bibfield  {title} {\bibinfo {title} {Magnetic skyrmion materials},\ }\href@noop {} {\bibfield  {journal} {\bibinfo  {journal} {Chemical Reviews}\ }\textbf {\bibinfo {volume} {121}},\ \bibinfo {pages} {2857} (\bibinfo {year} {2020})}\BibitemShut {NoStop}%
\bibitem [{\citenamefont {Kim}\ \emph {et~al.}(2014)\citenamefont {Kim}, \citenamefont {Kim},\ and\ \citenamefont {Sasaki}}]{kim2014boltzmann}%
  \BibitemOpen
  \bibfield  {author} {\bibinfo {author} {\bibfnamefont {K.-S.}\ \bibnamefont {Kim}}, \bibinfo {author} {\bibfnamefont {H.-J.}\ \bibnamefont {Kim}},\ and\ \bibinfo {author} {\bibfnamefont {M.}~\bibnamefont {Sasaki}},\ }\bibfield  {title} {\bibinfo {title} {Boltzmann equation approach to anomalous transport in a weyl metal},\ }\href@noop {} {\bibfield  {journal} {\bibinfo  {journal} {Physical Review B}\ }\textbf {\bibinfo {volume} {89}},\ \bibinfo {pages} {195137} (\bibinfo {year} {2014})}\BibitemShut {NoStop}%
\bibitem [{\citenamefont {Imran}\ and\ \citenamefont {Hershfield}(2018)}]{imran2018berry}%
  \BibitemOpen
  \bibfield  {author} {\bibinfo {author} {\bibfnamefont {M.}~\bibnamefont {Imran}}\ and\ \bibinfo {author} {\bibfnamefont {S.}~\bibnamefont {Hershfield}},\ }\bibfield  {title} {\bibinfo {title} {Berry curvature force and lorentz force comparison in the magnetotransport of weyl semimetals},\ }\href@noop {} {\bibfield  {journal} {\bibinfo  {journal} {Physical Review B}\ }\textbf {\bibinfo {volume} {98}},\ \bibinfo {pages} {205139} (\bibinfo {year} {2018})}\BibitemShut {NoStop}%
\bibitem [{\citenamefont {Zyuzin}(2017)}]{zyuzin2017magnetotransport}%
  \BibitemOpen
  \bibfield  {author} {\bibinfo {author} {\bibfnamefont {V.~A.}\ \bibnamefont {Zyuzin}},\ }\bibfield  {title} {\bibinfo {title} {Magnetotransport of weyl semimetals due to the chiral anomaly},\ }\href@noop {} {\bibfield  {journal} {\bibinfo  {journal} {Physical Review B}\ }\textbf {\bibinfo {volume} {95}},\ \bibinfo {pages} {245128} (\bibinfo {year} {2017})}\BibitemShut {NoStop}%
\bibitem [{\citenamefont {Das}\ and\ \citenamefont {Agarwal}(2019{\natexlab{b}})}]{das2019linear}%
  \BibitemOpen
  \bibfield  {author} {\bibinfo {author} {\bibfnamefont {K.}~\bibnamefont {Das}}\ and\ \bibinfo {author} {\bibfnamefont {A.}~\bibnamefont {Agarwal}},\ }\bibfield  {title} {\bibinfo {title} {Linear magnetochiral transport in tilted type-i and type-ii weyl semimetals},\ }\href@noop {} {\bibfield  {journal} {\bibinfo  {journal} {Physical Review B}\ }\textbf {\bibinfo {volume} {99}},\ \bibinfo {pages} {085405} (\bibinfo {year} {2019}{\natexlab{b}})}\BibitemShut {NoStop}%
\bibitem [{\citenamefont {Duval}\ \emph {et~al.}(2006)\citenamefont {Duval}, \citenamefont {Horv{\'a}th}, \citenamefont {Horvathy}, \citenamefont {Martina},\ and\ \citenamefont {Stichel}}]{duval2006comment}%
  \BibitemOpen
  \bibfield  {author} {\bibinfo {author} {\bibfnamefont {C.}~\bibnamefont {Duval}}, \bibinfo {author} {\bibfnamefont {Z.}~\bibnamefont {Horv{\'a}th}}, \bibinfo {author} {\bibfnamefont {P.~A.}\ \bibnamefont {Horvathy}}, \bibinfo {author} {\bibfnamefont {L.}~\bibnamefont {Martina}},\ and\ \bibinfo {author} {\bibfnamefont {P.}~\bibnamefont {Stichel}},\ }\bibfield  {title} {\bibinfo {title} {Comment on “berry phase correction to electron<? format?> density of states in solids”},\ }\href@noop {} {\bibfield  {journal} {\bibinfo  {journal} {Physical Review Letters}\ }\textbf {\bibinfo {volume} {96}},\ \bibinfo {pages} {099701} (\bibinfo {year} {2006})}\BibitemShut {NoStop}%
\bibitem [{\citenamefont {Bruus}\ and\ \citenamefont {Flensberg}(2004)}]{bruus2004many}%
  \BibitemOpen
  \bibfield  {author} {\bibinfo {author} {\bibfnamefont {H.}~\bibnamefont {Bruus}}\ and\ \bibinfo {author} {\bibfnamefont {K.}~\bibnamefont {Flensberg}},\ }\href@noop {} {\emph {\bibinfo {title} {Many-body quantum theory in condensed matter physics: an introduction}}}\ (\bibinfo  {publisher} {Oxford university press},\ \bibinfo {year} {2004})\BibitemShut {NoStop}%
\bibitem [{\citenamefont {Ziman}(1979)}]{ziman1979principles}%
  \BibitemOpen
  \bibfield  {author} {\bibinfo {author} {\bibfnamefont {J.~M.}\ \bibnamefont {Ziman}},\ }\href@noop {} {\emph {\bibinfo {title} {Principles of the Theory of Solids}}}\ (\bibinfo  {publisher} {Cambridge university press},\ \bibinfo {year} {1979})\BibitemShut {NoStop}%
\bibitem [{\citenamefont {Roy}\ \emph {et~al.}(2016)\citenamefont {Roy}, \citenamefont {Juri{\v{c}}i{\'c}},\ and\ \citenamefont {Das~Sarma}}]{roy2016universal}%
  \BibitemOpen
  \bibfield  {author} {\bibinfo {author} {\bibfnamefont {B.}~\bibnamefont {Roy}}, \bibinfo {author} {\bibfnamefont {V.}~\bibnamefont {Juri{\v{c}}i{\'c}}},\ and\ \bibinfo {author} {\bibfnamefont {S.}~\bibnamefont {Das~Sarma}},\ }\bibfield  {title} {\bibinfo {title} {Universal optical conductivity of a disordered weyl semimetal},\ }\href@noop {} {\bibfield  {journal} {\bibinfo  {journal} {Scientific reports}\ }\textbf {\bibinfo {volume} {6}},\ \bibinfo {pages} {32446} (\bibinfo {year} {2016})}\BibitemShut {NoStop}%
\bibitem [{\citenamefont {Ahmad}\ \emph {et~al.}(2026)\citenamefont {Ahmad}, \citenamefont {Bhalla}, \citenamefont {Nandy},\ and\ \citenamefont {Nag}}]{ahmad2026semiclassical}%
  \BibitemOpen
  \bibfield  {author} {\bibinfo {author} {\bibfnamefont {A.}~\bibnamefont {Ahmad}}, \bibinfo {author} {\bibfnamefont {P.}~\bibnamefont {Bhalla}}, \bibinfo {author} {\bibfnamefont {S.}~\bibnamefont {Nandy}},\ and\ \bibinfo {author} {\bibfnamefont {T.}~\bibnamefont {Nag}},\ }\bibfield  {title} {\bibinfo {title} {Semiclassical theory of frequency dependent linear magneto-optical transport in weyl semimetals},\ }\href@noop {} {\bibfield  {journal} {\bibinfo  {journal} {arXiv preprint arXiv:2604.11527}\ } (\bibinfo {year} {2026})}\BibitemShut {NoStop}%
\bibitem [{\citenamefont {Sharma}\ \emph {et~al.}(2023{\natexlab{a}})\citenamefont {Sharma}, \citenamefont {Nandy}, \citenamefont {Raman},\ and\ \citenamefont {Tewari}}]{PhysRevB.107.115161}%
  \BibitemOpen
  \bibfield  {author} {\bibinfo {author} {\bibfnamefont {G.}~\bibnamefont {Sharma}}, \bibinfo {author} {\bibfnamefont {S.}~\bibnamefont {Nandy}}, \bibinfo {author} {\bibfnamefont {K.~V.}\ \bibnamefont {Raman}},\ and\ \bibinfo {author} {\bibfnamefont {S.}~\bibnamefont {Tewari}},\ }\bibfield  {title} {\bibinfo {title} {Decoupling intranode and internode scattering in weyl fermions},\ }\href {https://doi.org/10.1103/PhysRevB.107.115161} {\bibfield  {journal} {\bibinfo  {journal} {Phys. Rev. B}\ }\textbf {\bibinfo {volume} {107}},\ \bibinfo {pages} {115161} (\bibinfo {year} {2023}{\natexlab{a}})}\BibitemShut {NoStop}%
\bibitem [{\citenamefont {Li}\ \emph {et~al.}(2021)\citenamefont {Li}, \citenamefont {Heinonen}, \citenamefont {Burkov},\ and\ \citenamefont {Zhang}}]{li2021nonlinear}%
  \BibitemOpen
  \bibfield  {author} {\bibinfo {author} {\bibfnamefont {R.-H.}\ \bibnamefont {Li}}, \bibinfo {author} {\bibfnamefont {O.~G.}\ \bibnamefont {Heinonen}}, \bibinfo {author} {\bibfnamefont {A.~A.}\ \bibnamefont {Burkov}},\ and\ \bibinfo {author} {\bibfnamefont {S.~S.-L.}\ \bibnamefont {Zhang}},\ }\bibfield  {title} {\bibinfo {title} {Nonlinear hall effect in weyl semimetals induced by chiral anomaly},\ }\href@noop {} {\bibfield  {journal} {\bibinfo  {journal} {Physical Review B}\ }\textbf {\bibinfo {volume} {103}},\ \bibinfo {pages} {045105} (\bibinfo {year} {2021})}\BibitemShut {NoStop}%
\bibitem [{\citenamefont {Zeng}\ \emph {et~al.}(2022)\citenamefont {Zeng}, \citenamefont {Nandy},\ and\ \citenamefont {Tewari}}]{zeng2022chiral}%
  \BibitemOpen
  \bibfield  {author} {\bibinfo {author} {\bibfnamefont {C.}~\bibnamefont {Zeng}}, \bibinfo {author} {\bibfnamefont {S.}~\bibnamefont {Nandy}},\ and\ \bibinfo {author} {\bibfnamefont {S.}~\bibnamefont {Tewari}},\ }\bibfield  {title} {\bibinfo {title} {Chiral anomaly induced nonlinear nernst and thermal hall effects in weyl semimetals},\ }\href@noop {} {\bibfield  {journal} {\bibinfo  {journal} {Physical Review B}\ }\textbf {\bibinfo {volume} {105}},\ \bibinfo {pages} {125131} (\bibinfo {year} {2022})}\BibitemShut {NoStop}%
\bibitem [{\citenamefont {Kharzeev}(2014)}]{kharzeev2014chiral}%
  \BibitemOpen
  \bibfield  {author} {\bibinfo {author} {\bibfnamefont {D.~E.}\ \bibnamefont {Kharzeev}},\ }\bibfield  {title} {\bibinfo {title} {The chiral magnetic effect and anomaly-induced transport},\ }\href@noop {} {\bibfield  {journal} {\bibinfo  {journal} {Progress in Particle and Nuclear Physics}\ }\textbf {\bibinfo {volume} {75}},\ \bibinfo {pages} {133} (\bibinfo {year} {2014})}\BibitemShut {NoStop}%
\bibitem [{\citenamefont {Kumar}\ \emph {et~al.}(2022)\citenamefont {Kumar}, \citenamefont {Nagpal}, \citenamefont {Sudesh},\ and\ \citenamefont {Patnaik}}]{kumar2022chiral}%
  \BibitemOpen
  \bibfield  {author} {\bibinfo {author} {\bibfnamefont {P.}~\bibnamefont {Kumar}}, \bibinfo {author} {\bibfnamefont {V.}~\bibnamefont {Nagpal}}, \bibinfo {author} {\bibnamefont {Sudesh}},\ and\ \bibinfo {author} {\bibfnamefont {S.}~\bibnamefont {Patnaik}},\ }\bibfield  {title} {\bibinfo {title} {Chiral anomaly induced negative magnetoresistance and weak anti-localization in weyl semimetal bi0. 97sb0. 03 alloy},\ }\href@noop {} {\bibfield  {journal} {\bibinfo  {journal} {Journal of Physics: Condensed Matter}\ }\textbf {\bibinfo {volume} {34}},\ \bibinfo {pages} {055601} (\bibinfo {year} {2022})}\BibitemShut {NoStop}%
\bibitem [{\citenamefont {Johansson}\ \emph {et~al.}(2019)\citenamefont {Johansson}, \citenamefont {Henk},\ and\ \citenamefont {Mertig}}]{johansson2019chiral}%
  \BibitemOpen
  \bibfield  {author} {\bibinfo {author} {\bibfnamefont {A.}~\bibnamefont {Johansson}}, \bibinfo {author} {\bibfnamefont {J.}~\bibnamefont {Henk}},\ and\ \bibinfo {author} {\bibfnamefont {I.}~\bibnamefont {Mertig}},\ }\bibfield  {title} {\bibinfo {title} {Chiral anomaly in type-i weyl semimetals: Comprehensive analysis within a semiclassical fermi surface harmonics approach},\ }\href@noop {} {\bibfield  {journal} {\bibinfo  {journal} {Physical Review B}\ }\textbf {\bibinfo {volume} {99}},\ \bibinfo {pages} {075114} (\bibinfo {year} {2019})}\BibitemShut {NoStop}%
\bibitem [{\citenamefont {Ahmad}(2026)}]{ahmad2026chiral}%
  \BibitemOpen
  \bibfield  {author} {\bibinfo {author} {\bibfnamefont {A.}~\bibnamefont {Ahmad}},\ }\bibfield  {title} {\bibinfo {title} {Chiral anomaly and planar hall conductance in pseudospin-$1 $ fermions},\ }\href@noop {} {\bibfield  {journal} {\bibinfo  {journal} {arXiv preprint arXiv:2606.11364}\ } (\bibinfo {year} {2026})}\BibitemShut {NoStop}%
\bibitem [{\citenamefont {Mahan}(2008)}]{mahan20089}%
  \BibitemOpen
  \bibfield  {author} {\bibinfo {author} {\bibfnamefont {G.~D.}\ \bibnamefont {Mahan}},\ }\href@noop {} {\emph {\bibinfo {title} {9. Many-Particle Systems}}}\ (\bibinfo  {publisher} {Princeton University Press},\ \bibinfo {year} {2008})\BibitemShut {NoStop}%
\bibitem [{\citenamefont {Abers}(2004)}]{abers2004quantum}%
  \BibitemOpen
  \bibfield  {author} {\bibinfo {author} {\bibfnamefont {E.}~\bibnamefont {Abers}},\ }\href {https://books.google.co.in/books?id=_aBkQgAACAAJ} {\emph {\bibinfo {title} {Quantum Mechanics}}}\ (\bibinfo  {publisher} {Pearson Education},\ \bibinfo {year} {2004})\BibitemShut {NoStop}%
\bibitem [{\citenamefont {Morimoto}\ \emph {et~al.}(2016)\citenamefont {Morimoto}, \citenamefont {Zhong}, \citenamefont {Orenstein},\ and\ \citenamefont {Moore}}]{morimoto2016semiclassical}%
  \BibitemOpen
  \bibfield  {author} {\bibinfo {author} {\bibfnamefont {T.}~\bibnamefont {Morimoto}}, \bibinfo {author} {\bibfnamefont {S.}~\bibnamefont {Zhong}}, \bibinfo {author} {\bibfnamefont {J.}~\bibnamefont {Orenstein}},\ and\ \bibinfo {author} {\bibfnamefont {J.~E.}\ \bibnamefont {Moore}},\ }\bibfield  {title} {\bibinfo {title} {Semiclassical theory of nonlinear magneto-optical responses with applications to topological dirac/weyl semimetals},\ }\href@noop {} {\bibfield  {journal} {\bibinfo  {journal} {Physical Review B}\ }\textbf {\bibinfo {volume} {94}},\ \bibinfo {pages} {245121} (\bibinfo {year} {2016})}\BibitemShut {NoStop}%
\bibitem [{\citenamefont {Sodemann}\ and\ \citenamefont {Fu}(2015)}]{sodemann2015quantum}%
  \BibitemOpen
  \bibfield  {author} {\bibinfo {author} {\bibfnamefont {I.}~\bibnamefont {Sodemann}}\ and\ \bibinfo {author} {\bibfnamefont {L.}~\bibnamefont {Fu}},\ }\bibfield  {title} {\bibinfo {title} {Quantum nonlinear hall effect induced by berry curvature dipole in time-reversal invariant materials},\ }\href@noop {} {\bibfield  {journal} {\bibinfo  {journal} {Physical review letters}\ }\textbf {\bibinfo {volume} {115}},\ \bibinfo {pages} {216806} (\bibinfo {year} {2015})}\BibitemShut {NoStop}%
\bibitem [{\citenamefont {Dagnino}\ \emph {et~al.}(2025)\citenamefont {Dagnino}, \citenamefont {Liu},\ and\ \citenamefont {Neupert}}]{dagnino2025non}%
  \BibitemOpen
  \bibfield  {author} {\bibinfo {author} {\bibfnamefont {A.~K.}\ \bibnamefont {Dagnino}}, \bibinfo {author} {\bibfnamefont {X.}~\bibnamefont {Liu}},\ and\ \bibinfo {author} {\bibfnamefont {T.}~\bibnamefont {Neupert}},\ }\bibfield  {title} {\bibinfo {title} {Non-linear transport in multifold semimetals},\ }\href@noop {} {\bibfield  {journal} {\bibinfo  {journal} {arXiv preprint arXiv:2512.08891}\ } (\bibinfo {year} {2025})}\BibitemShut {NoStop}%
\bibitem [{\citenamefont {Son}\ and\ \citenamefont {Spivak}(2013)}]{son2013chiral}%
  \BibitemOpen
  \bibfield  {author} {\bibinfo {author} {\bibfnamefont {D.}~\bibnamefont {Son}}\ and\ \bibinfo {author} {\bibfnamefont {B.}~\bibnamefont {Spivak}},\ }\bibfield  {title} {\bibinfo {title} {Chiral anomaly and classical negative magnetoresistance of weyl metals},\ }\href@noop {} {\bibfield  {journal} {\bibinfo  {journal} {Physical Review B}\ }\textbf {\bibinfo {volume} {88}},\ \bibinfo {pages} {104412} (\bibinfo {year} {2013})}\BibitemShut {NoStop}%
\bibitem [{\citenamefont {Varma~K}\ \emph {et~al.}(2024)\citenamefont {Varma~K}, \citenamefont {Ahmad}, \citenamefont {Tewari},\ and\ \citenamefont {Sharma}}]{varma2024magnetotransport}%
  \BibitemOpen
  \bibfield  {author} {\bibinfo {author} {\bibfnamefont {G.}~\bibnamefont {Varma~K}}, \bibinfo {author} {\bibfnamefont {A.}~\bibnamefont {Ahmad}}, \bibinfo {author} {\bibfnamefont {S.}~\bibnamefont {Tewari}},\ and\ \bibinfo {author} {\bibfnamefont {G.}~\bibnamefont {Sharma}},\ }\bibfield  {title} {\bibinfo {title} {Magnetotransport in spin-orbit coupled noncentrosymmetric and weyl metals},\ }\href@noop {} {\bibfield  {journal} {\bibinfo  {journal} {Physical Review B}\ }\textbf {\bibinfo {volume} {109}},\ \bibinfo {pages} {165114} (\bibinfo {year} {2024})}\BibitemShut {NoStop}%
\bibitem [{\citenamefont {Sharma}\ \emph {et~al.}(2023{\natexlab{b}})\citenamefont {Sharma}, \citenamefont {Nandy}, \citenamefont {Raman},\ and\ \citenamefont {Tewari}}]{sharma2023decoupling}%
  \BibitemOpen
  \bibfield  {author} {\bibinfo {author} {\bibfnamefont {G.}~\bibnamefont {Sharma}}, \bibinfo {author} {\bibfnamefont {S.}~\bibnamefont {Nandy}}, \bibinfo {author} {\bibfnamefont {K.~V.}\ \bibnamefont {Raman}},\ and\ \bibinfo {author} {\bibfnamefont {S.}~\bibnamefont {Tewari}},\ }\bibfield  {title} {\bibinfo {title} {Decoupling intranode and internode scattering in weyl fermions},\ }\href@noop {} {\bibfield  {journal} {\bibinfo  {journal} {Physical Review B}\ }\textbf {\bibinfo {volume} {107}},\ \bibinfo {pages} {115161} (\bibinfo {year} {2023}{\natexlab{b}})}\BibitemShut {NoStop}%
\bibitem [{\citenamefont {Varma~K}\ \emph {et~al.}(2026)\citenamefont {Varma~K}, \citenamefont {Ahmad},\ and\ \citenamefont {Sharma}}]{varma2026strain}%
  \BibitemOpen
  \bibfield  {author} {\bibinfo {author} {\bibfnamefont {G.}~\bibnamefont {Varma~K}}, \bibinfo {author} {\bibfnamefont {A.}~\bibnamefont {Ahmad}},\ and\ \bibinfo {author} {\bibfnamefont {G.}~\bibnamefont {Sharma}},\ }\bibfield  {title} {\bibinfo {title} {Strain-induced ettingshausen effect in spin-orbit coupled noncentrosymmetric metals},\ }\href@noop {} {\bibfield  {journal} {\bibinfo  {journal} {Physical Review B}\ }\textbf {\bibinfo {volume} {113}},\ \bibinfo {pages} {075118} (\bibinfo {year} {2026})}\BibitemShut {NoStop}%
\end{thebibliography}%
\end{document}